\documentclass[a4paper,11pt]{article}
\pdfoutput=1 

\usepackage{jcappub} 
\usepackage[T1]{fontenc}

\usepackage{color}
\usepackage{comment}
\usepackage{multirow}
\usepackage{float}
\usepackage{subcaption}
\usepackage{booktabs}
\usepackage{aas_macros}
\usepackage{tabularx}

\title{\boldmath The Galactic Dynamics of Free-Floating Planets: From Ejection Kicks to Microlensing}

\author[1]{Zara Sayed,}
\author[1]{Stefano Profumo,}
\author[2,3]{and Nolan Smyth}

\affiliation[1]{Department of Physics and Santa Cruz Institute for Particle Physics,\\
University of California, Santa Cruz, CA 95064, USA}
\affiliation[2]{Department of Physics, Universit\'e de Montr\'eal, Montr\'eal, QC, Canada}
\affiliation[3]{Mila -- Quebec Artificial Intelligence Institute, Montr\'eal, QC, Canada}

\emailAdd{zsayed@ucsc.edu}

\abstract{Free-floating planets (FFPs) may retain dynamical signatures of the mechanisms that eject them from their host systems. We integrate $10^6$ collisionless FFP test particles for $10^8$~yr in a static, phenomenological Galactic potential, comparing a mass-independent kick model with a mass-coupled prescription ($\sigma_{\rm CB}\propto M^{-0.15}$, $v_{\rm PL}\propto M^{-0.5}$) against a matched no-kick control. The imposed mass dependence is recovered at injection, with the combined-channel median coupled kick decreasing from $10.716$ to $0.708~{\rm km\,s^{-1}}$ across five mass bins---a factor of $\sim15.1$. A mass-dependent trend remains visible after $100$~Myr of Galactic propagation in the matched kicked-versus-control displacement, whose median decreases from $0.7001$ to $0.0539$~kpc across the same bins. The remaining numerical distinction is concentrated in this mass-resolved differential displacement rather than in the bulk phase-space moments: the final velocity dispersions of the null and coupled kicked populations are nearly identical, only a few percent of particles leave the adopted disk region, and the fraction of particles satisfying the formal Galactic-unbound criterion ($E_i>0$) remains negligible in every diagnostic we report. Comparing each kicked population against its matched no-kick control shows that much of the overall kinematic heating arises from the evolution of the initially warm, nonequilibrium disk rather than from the ejection kick itself, isolating the kick's smaller contribution. A supplementary $1$\,Gyr integration suggests that the late-time evolution does not erase the qualitative pattern, although a definitive long-term assessment requires higher-resolution integrations and more dynamically self-consistent initial conditions. We use the resulting phase-space distribution to construct a shape-level microlensing event-rate calculation for the Roman Space Telescope, which predicts the relative timescale distribution for FFPs in each mass bin. Because the calculation is not tied to an absolute FFP density or survey selection function, it should be interpreted as a prediction of relative shape rather than absolute event yield.}


\begin{document}
\maketitle
\flushbottom

\section{Introduction}

The discovery of FFPs - planetary-mass objects unbound to any host star - represents one of the most intriguing developments in exoplanet science. Microlensing surveys such as MOA and OGLE have provided tantalizing evidence that FFPs may be common in the Galaxy, with early results suggesting that they could rival the stellar population in abundance \cite{Sumi2011}. Subsequent analyses, however, have painted a more nuanced picture, indicating that Jupiter-mass FFPs are relatively rare while lower-mass FFPs remain plausible in large numbers \cite{Mróz2017,Mróz2018,Mróz2019}.

Several dynamical channels can contribute to the production of FFPs. Planet-planet scattering in unstable multi-planet systems naturally leads to ejections on Myr timescales, especially in systems hosting multiple giant planets like Neptune and Jupiter mass \cite{RasioFord1996,Chatterjee2008,VerasRaymond2012,BhaskarPerets2025}. Stellar flybys in clusters can unbind wide-orbit planets and planetesimals, producing a population of unbound bodies with velocities tied to the cluster environment \cite{Pfalzner2021,YuLai2024}. Circumbinary planets are particularly vulnerable to dynamical instabilities and can be expelled at significant velocities \cite{Coleman2024}. Additional channels include post-main-sequence stellar evolution, where planets can be destabilized and lost during mass loss phases \cite{Veras2016}.

These processes imprint distinct ejection velocity distributions, typically peaked at a few km\,s$^{-1}$ but with non-negligible high-velocity tails \citep{SutherlandFabrycky2016,BhaskarPerets2025,ZhaiLee2025}. The subsequent dynamical evolution of ejected planets in the Galaxy depends on their initial phase-space distribution. Once unbound, FFPs behave dynamically like collisionless test particles in the Galactic potential \citep{BinneyTremaine2008}. A key question is therefore whether a mass dependence in the initial ejection velocity can survive subsequent Galactic propagation and remain detectable in the spatial or velocity distribution. We note, however, that the warm-disk initial condition adopted here is a Jeans-like, locally motivated construction rather than a fully self-consistent equilibrium distribution function for the adopted triaxial potential. The resulting evolution must therefore be interpreted with care, since changes in the phase-space distribution can reflect both the imposed ejection kicks and the subsequent evolution of the initial non-equilibrium state.

Observationally, microlensing remains the most powerful tool to detect FFPs. The short timescales of FFP lensing events, combined with their absence of detectable stellar hosts, distinguish them from stellar lenses \citep{Paczynski1986}. Current constraints from OGLE and MOA point to a population of FFPs smaller than initially claimed but still potentially significant \citep{Mróz2017,Mróz2018}. The forthcoming Nancy Grace Roman Space Telescope will provide a transformative dataset, monitoring the Galactic bulge with unparalleled cadence and sensitivity \citep{Penny2019,Johnson2020}. Roman will detect large samples of FFP microlensing events, down to planetary masses comparable to Earth and below, enabling a direct measurement of their abundance and distribution \citep{Chachan_2024}. Accurately predicting Roman's yield requires a robust theoretical framework for the phase-space distribution of FFPs.

Previous cluster-scale studies have demonstrated that FFP ejection velocities influence whether planets remain associated with their parent cluster, and have characterized the resulting velocity distributions and cluster-escape and retention timescales \citep{2015MNRAS.449.3543W,2001MNRAS.322..859B, ColemanDeRocco2025}. Simulations have begun to model the Galactic FFP population from realistic planet-formation and ejection histories, including mass-dependent ejection locations and velocities. Although these studies have investigated FFP ejection velocities, cluster retention, and the resulting Galactic FFP population, the mapping between an explicit mass-dependent ejection-velocity prescription and the subsequent spatial redistribution of FFPs in a Galactic potential remains comparatively less explored. Our calculation takes the next dynamical step by propagating the resulting FFP population in a fixed Galactic potential and examining how a mass-dependent kick prescription maps into spatial redistribution on Myr--100~Myr timescales; this is a controlled dynamical experiment comparing mass-independent and explicitly mass-coupled kick prescriptions using matched initial conditions and a no-kick control, rather than a claim that the mass dependence of FFP spatial distributions has not previously been considered.

In this work, we combine sampling of initial conditions with a collisionless test-particle integration \citep{2024MNRAS.527..414C} of free-floating planets in the Milky Way, using a matched no-kick control population as the central diagnostic tool. Schematically, the calculation proceeds along the chain
\begin{equation}
M_{\rm FFP} \;\rightarrow\; v_{\rm kick}(M)
\;\rightarrow\; \Delta r(M,t)
\;\rightarrow\; P(R,z\mid M,t),
\end{equation}
\label{eq:schematic_chain}
and the central question is whether a mass-dependent ejection kick imposed at birth produces a measurable mass-dependent signature in the resulting spatial and velocity distribution after $10^8$~yr of Galactic propagation. The kicked and control populations are initialized with identical positions, masses, and pre-kick velocities, with the control receiving no ejection kick. This matched construction allows the subsequent evolution of the control population to be separated from the incremental effect of the kick. We first establish the mass dependence of the injected kick distribution, then examine its propagated spatial and kinematic signatures, and finally connect the resulting phase-space distribution to a schematic microlensing calculation for the Nancy Grace Roman Space Telescope. 

The remainder of the paper is organized as follows. Section~\ref{sec:ejection_mechanisms} reviews the physical channels responsible for FFP ejection and their relative importance as a function of planet mass. Section~\ref{sec:kicks} develops the phenomenological source term and kick-velocity prescription used to sample ejection velocities for each channel. Section~\ref{sec:numerical_setup} describes the numerical setup, including the initial spatial and kinematic sampling, the disk--bulge--halo Galactic potential, and the leapfrog integration scheme. Section~\ref{sec:diagnostics} defines the propagation, net-displacement, and mass-resolved diagnostics used to characterize the evolved population, including the matched no-kick control. Section~\ref{sec:results} presents the numerical results: disk-departure and velocity-evolution diagnostics, the mass dependence recovered at injection, and the propagation-distance, phase-space, and kicked-versus-control comparisons that establish the surviving mass-dependent spatial response and the importance of the non-equilibrium control evolution. Section~\ref{sec:Microlensing} constructs an approximate phase-space distribution function from the simulation output and formulates the corresponding microlensing event-rate formalism for Roman, and Sec.~\ref{sec:numerical_implementation} reports the resulting schematic event-rate predictions Section~\ref{sec:conclusion}. Appendix~\ref{app:parameters} collects, for reference and reproducibility, the full set of adopted parameter values, sampling prescriptions, and numerical settings underlying the simulations, together with the supplementary $N=10^4$, 1~Gyr integration.

\section{Ejection Mechanisms and Kick Velocity Distributions}
\label{sec:ejection_mechanisms}

FFPs come through a variety of ejection mechanisms, each imprinting distinct signatures on their initial velocity distributions \citep{BhaskarPerets2025}. Understanding these channels is important for interpreting the present-day phase-space structure of the Galactic FFP population and for motivating physically plausible ejection scenarios.

One prominent formation pathway is \emph{planet-planet scattering}, in which close gravitational encounters within a planetary system destabilize orbits and lead to the ejection of one or more planets \citep{BhaskarPerets2025}. In systems hosting massive, Jupiter-like planets, smaller bodies can receive strong gravitational kicks during close encounters, resulting in ejection velocities of several kilometers per second. Ejection velocities in planet-planet scattering depend on the initial distance of the ejected planet from the host star \citep{BhaskarPerets2025}. Such scattering-induced ejections are expected to dominate in dynamically active multi-planet systems and naturally produce a broad range of kick speeds \citep{ColemanDeRocco2025,ZhaiLee2025}.

FFPs are also produced following \emph{stellar mass loss during supernova explosions}. When a massive star undergoes a supernova, the sudden loss of a large fraction of its mass dramatically weakens the gravitational potential binding its planets. As a result, previously bound planets become unbound and drift freely through the Galaxy. The resulting ejection velocities depend on the pre-explosion orbital configuration and the magnitude of the mass loss, and this produces high-velocity FFPs and compact remnants \citep{2026PASP..138h2001R}.

An alternative pathway involves \emph{star-like formation}, in which planetary-mass objects form directly through the gravitational collapse of gas and dust in molecular clouds. In this case, the objects never orbit a host star and instead form as isolated bodies. These objects are expected to inherit relatively low velocities characteristic of their natal star-forming environments rather than impulsive ejection kicks, similar to the local stellar velocity dispersion \citep{2022NatAs...6...89M, albrow2025ejectionvelocitiesinterstellarobjects}.

\emph{Stellar flybys} provide an additional mechanism for liberating planets from their host systems. Close encounters between stars can perturb planetary orbits, particularly in dense star-forming regions where stellar interactions are frequent; the rates and velocity distributions depend on cluster density and encounter parameters \citep{YuLai2024,2015MNRAS.449.3543W}. Such encounters preferentially destabilize wide-orbit, low-mass bodies, with resulting ejection velocities of order $0.5$--$2\,\mathrm{km\,s^{-1}}$ \citep{Pfalzner2021}, distinctly lower than the $\sim2$--$6\,\mathrm{km\,s^{-1}}$ typical of planet--planet scattering (Sec.~\ref{sec:kicks}).

\emph{Binary host systems} are capable of producing significantly higher ejection speeds than single-star systems due to their deeper and time-varying gravitational potentials \cite{Coleman2024, boffin2024importancebinarystars}. 

These mechanisms motivate caution in interpreting a single mixture of CB and PL particles as \emph{the} astrophysical FFP population. Following Sec.~\ref{sec:kicks}, our two modeled channels are therefore analyzed separately throughout. 

Any apparent mass trend in the ALL sample can reflect the changing CB:PL mixture fraction with mass as much as the underlying kick physics, since the two channels have different characteristic kick amplitudes ($\sigma_{\rm CB}=10$ vs.\ $v_k=4~{\rm km\,s^{-1}}$). The CB and PL channel-resolved results (Tables~\ref{tab:mass_binned_kicks}, \ref{tab:control_comparison}) are the primary physical results of this paper.

\subsection{Relative Importance of Ejection Channels as a Function of Planet Mass}
\label{sec:Importance_eject}

The importance of different ejection mechanisms depends strongly on planet mass. Planet--planet scattering is the dominant ejection channel for giant planets, as gravitational instabilities in multi-planet systems efficiently eject massive bodies once gas disk damping disappears; bound planets tend to be more massive than ejected ones in many simulations \citep{BhaskarPerets2025}. Lower-mass planets experience weaker mutual scattering and are therefore less likely to be ejected through this mechanism alone \citep{ZhaiLee2025,Chachan_2024}. Existing theoretical work suggests that giant FFPs are dominated by planet--planet scattering, while lower-mass free floaters may receive significant contributions from external perturbations such as stellar encounters; however, quantitative predictions remain uncertain \citep{guo2025formationfreefloatingplanetsejection}.

Here, we use a coupled kick scaling, and the ejection channels discussed in Section~\ref{sec:ejection_mechanisms} are not modeled individually. Instead, we adopt a phenomenological sampling distribution that provides a unified statistical description of the FFP population by combining the mass distribution and characteristic kick-velocity distributions associated with the dominant ejection mechanisms. The complete formulation of this injection distribution is presented in Section~\ref{sec:kicks} and captures the FFP population's overall dynamical properties through the parameters introduced there.

\section{Kick Velocity Distribution Model}
\label{sec:kicks}

Building on the ejection channels and their mass dependence reviewed in
Section~\ref{sec:ejection_mechanisms}, we now introduce the central theoretical ingredient of this work: a physically motivated kick-velocity model that represents the collective effects of the various ejection channels.

$N$-body simulations consistently show that the velocity distribution of planets ejected from planetary systems exhibits a pronounced low-velocity peak together with a rapidly declining high-velocity tail
\cite{ColemanDeRocco2025,2016MNRAS.461L.107M, Pfalzner2021}. Several studies also suggest that the ejection velocity may depend on planet mass. Motivated by these results, we adopt the phenomenological kick model described in Section~\ref{sec:Source_term}.

\subsection{Phenomenological Source Term}
\label{sec:Source_term}

To model the population of FFPs produced by circumbinary systems, we adopt the factorized source term

\begin{equation}
\frac{dN_{\rm ej}}{dM\,dv}
=
A\,M^{-\alpha}
\left[1+\left(\frac{M}{M_b}\right)^\gamma\right]^{-1}
\sqrt{\frac{2}{\pi}}\,
\frac{v^2}{\sigma_{\rm CB}^3}
\exp\!\left(-\frac{v^2}{2\sigma_{\rm CB}^2}\right).
\end{equation}
This expression is intended as a phenomenological ansatz rather than a unique prediction of any individual simulation suite.

The factor $A M^{-\alpha}$ represents the underlying planetary mass spectrum produced in circumbinary systems. Power-law mass functions are widely used in exoplanet demographic studies, with giant-planet occurrence rates commonly described by $dN/dM \propto M^{-\alpha}$ and $\alpha$ of order unity \cite{Cumming2008}. In the absence of a well-constrained mass spectrum for FFPs originating from circumbinary systems, this provides a simple and physically motivated parametrization.

The suppression factor

\[
\left[1+\left(\frac{M}{M_b}\right)^\gamma\right]^{-1}
\]

encodes the expectation that lower-mass planets are more easily ejected, while planets with masses comparable to the dominant perturbers are harder to expel. Existing circumbinary formation and scattering studies support this qualitative picture, in that unstable planets are frequently lost by ejection and the surviving bound planets tend to be preferentially more massive. We emphasize, however, that this particular functional form is introduced here as a smooth interpolation, not as a standard analytic law established in the literature \cite{Coleman2024,ColemanDeRocco2025}. 

The parameter $\gamma$ controls the sharpness of the transition between the efficiently ejected low-mass regime and the increasingly suppressed high-mass regime. Larger values of $\gamma$ produce a sharper turnover near $M \sim M_b$, whereas smaller values yield a more gradual transition. In the absence of direct constraints from simulations, we adopt $\gamma \sim 1$--$3$, with a fiducial value of $\gamma=2$.

The parameter $M_b$ is treated as a phenomenological turnover mass rather than a directly measured quantity. We adopt

\[
M_b \sim 1\text{--}10\,M_\oplus,
\]

with a fiducial value of $M_b=3\,M_\oplus$. This value is motivated by numerical studies suggesting that circumbinary systems efficiently produce FFPs above approximately Earth mass, while the ejected population becomes increasingly suppressed at giant-planet masses \cite{Coleman2024,ColemanDeRocco2025}.

The excess-speed distribution is modeled using a Maxwellian,

\[
p(v)=
\sqrt{\frac{2}{\pi}}\,
\frac{v^2}{\sigma_{\rm CB}^3}
\exp\!\left(-\frac{v^2}{2\sigma_{\rm CB}^2}\right),
\]

where $\sigma_{\rm CB}$ is the characteristic kick velocity. Current circumbinary ejection simulations primarily identify a characteristic velocity scale, with typical excess speeds of order $\sim10~{\rm km\,s^{-1}}$, rather than a universal higher-order fitting function \cite{Coleman2024,ColemanDeRocco2025}. A Maxwellian therefore provides a simple one-parameter representation that can be readily propagated through Galactic dynamical calculations and microlensing rate integrals.

For the circumbinary source term we adopt

\[
\sigma_{\rm CB}\sim8\text{--}16~{\rm km\,s^{-1}},
\]

with a fiducial value of $\sigma_{\rm CB}=10~{\rm km\,s^{-1}}$, consistent with circumbinary ejection simulations \cite{Coleman2024,ColemanDeRocco2025}. The planetary mass spectrum is taken to follow

\[
\frac{dN_{\rm form}^{\rm CB}}{dM}\propto M^{-\alpha},
\]

where $\alpha\simeq1.3$ is adopted as the fiducial value, with a plausible range of $\alpha\sim1.0$--$1.5$ motivated by giant-planet occurrence studies \cite{Cumming2008,Fernandes2019}. Since the overall normalization is not directly constrained observationally, it is defined through

\[
\eta_{\rm CB}=\int_{M_{\min}}^{M_{\max}} dM\,A\,M^{-\alpha},
\]

where $\eta_{\rm CB}$ denotes the integrated abundance over the adopted mass interval. We adopt the broad phenomenological range $\eta_{\rm CB}\sim0.1$--$0.5$ planets per binary, motivated by observed circumbinary occurrence rates \cite{Armstrong2014}.

For comparison, we also consider a constant-kick model,

\[
\frac{dN_{\rm ej}}{dM\,dv}
=
A\,M^{-\alpha}\,\delta(v-v_k),
\]

where the delta function is replaced by a narrow Gaussian during numerical implementation. Recent $N$-body simulations of planet--planet scattering around single stars predict characteristic excess velocities of approximately $2$--$6~{\rm km\,s^{-1}}$, and we adopt the representative value $v_k=4~{\rm km\,s^{-1}}$ \cite{BhaskarPerets2025}. The mass spectrum again follows $dN/dM\propto M^{-\alpha}$ with $\alpha\simeq1.3$, while the normalization is defined through

\[
\eta_{\rm ej}=\int_{M_{\min}}^{M_{\max}} dM\,A\,M^{-\alpha},
\]

where $\eta_{\rm ej}$ represents the total number of ejected planets per star within the adopted mass interval. Based on recent scattering simulations, we adopt $\eta_{\rm ej}\sim2$--$8$ planets per star, corresponding to ejection fractions of approximately $40$--$80\%$ from planetary systems containing roughly $5$--$10$ planets \cite{BhaskarPerets2025}.

The primary goal of this work is to investigate how the Galactic distribution of FFPs responds to planet-mass-dependent ejection velocities and to quantify the resulting velocity and spatial distributions after post-ejection Galactic propagation. The ejection kick is applied at the initial time, $t=0$; the simulation therefore does not model an intrinsic distribution of planetary-system ejection times.

In the factorized source model adopted here, the planetary mass spectrum and kick-velocity distribution are treated as independent quantities. Schematically, we assume
\begin{equation}
\frac{dN}{dM\,d^3v}
\propto
\psi(M)\,f(v).
\label{eq:factorized_source}
\end{equation}

\subsection{Kick Prescription} 
\label{sec:kick_initialization} 

The simulation considers two prescriptions for the relation between FFP mass and ejection speed \citep{Coleman2024,ColemanDeRocco2025,BhaskarPerets2025}. In the mass-independent null model, the kick distribution is independent of mass within each channel. The CB channel is assigned three independent Gaussian velocity components with standard deviation $\sigma_{\rm CB}=10~{\rm km\,s^{-1}}$, and the resulting kick speed is the magnitude of this three-dimensional vector. The PL channel is assigned a narrow Gaussian distribution in kick speed centered on $v_k=4~{\rm km\,s^{-1}}$ with width $1~{\rm km\,s^{-1}}$~\citep{HuangLai2026}. In the mass-coupled model, the characteristic kick scale varies with planet mass. For the CB channel, 
\begin{equation} 
\sigma_{\rm CB,eff}(M) = \sigma_{\rm CB} \left(\frac{M}{M_\oplus}\right)^{-0.15}, 
\label{eq:cb_mass_coupling} 
\end{equation} 
while for the PL channel, 
\begin{equation}
v_{\rm PL,base}(M) = v_k \left(\frac{M}{M_\oplus}\right)^{-1/2}. \label{eq:pl_mass_coupling} 
\end{equation} 

The PL exponent of $-1/2$ is motivated by an approximate momentum-conservation-like scaling expected for scattering-driven ejections, $v\propto M^{-1/2}$, broadly consistent with the trends discussed by \citep{HuangLai2026} and \citep{BhaskarPerets2025}. The CB exponent of $-0.15$ is comparatively weaker and, unlike the PL exponent, is not derived from a specific analytic scaling; we adopt it as an illustrative empirical choice motivated qualitatively by the circumbinary ejection simulations of \citep{Coleman2024} and \citep{ColemanDeRocco2025}, and we note that this is explicitly a modeling choice that would benefit from a firmer physical or numerical basis in future work.

\section{Numerical Simulations Setup}
\label{sec:numerical_setup}

The Galactic potential is modeled as the sum of a \textbf{disk, triaxial bulge, and NFW-like dark-matter halo}, with accelerations calculated numerically using centered finite differences with a spatial step of $10^{-2}$~kpc. FFPs are initially sampled throughout the disk in radius, azimuth, and height, and their velocities are initialized using the local circular velocity plus a finite velocity dispersion in the revised warm-disk setup. Each particle then receives a single isotropic ejection kick according to its assigned mass and ejection channel, after which it is integrated as a test particle in the fixed Galactic potential. A matched \textbf{no-kick control population} is evolved alongside the kicked population to distinguish genuine ejection-induced kinematic changes from dynamical evolution of the initial distribution.

In short: the dynamical chain is
\begin{equation}
\Phi_{\rm disk}
+
\Phi_{\rm bulge}
+
\Phi_{\rm halo}
\rightarrow
\mathbf{a}
=
-\nabla\Phi
\rightarrow
v_{\rm circ}\;(V_0)
\rightarrow
\mathbf{v}_\star
\rightarrow
\mathbf{v}_{\rm kick}(M)
\rightarrow
\mathbf{v}_{\rm FFP}
\rightarrow
\text{leapfrog orbit integration}.
\end{equation}

\subsection{Initial Spatial and Kinematic Conditions} 
\label{sec:InitialConditions} 

The initial spatial and kinematic conditions describe the starting positions and motions assigned to each free-floating planet before its ejection kick. The FFP masses are sampled independently according to the prescribed mass functions for the two ejection channels: the CB population follows 

\begin{equation}
dN/dM\propto M^{-1.3} [1+(M/3M_\oplus)^2]^{-1}, 
\end{equation}

while the PL population follows 

\begin{equation}
dN/dM\propto M^{-1.3}. 
\end{equation}

The mass--velocity relationship is then introduced through the kick prescription: in the coupled model, CB kick dispersion scales as $\sigma_{\rm kick}\propto M^{-0.15}$, while PL kick speed scales as $v_{\rm kick}\propto M^{-0.5}$, so lower-mass FFPs receive systematically larger kicks. 

The kick velocity then depends on mass:

\begin{equation}
\sigma_{\rm CB}(M)
=
10
\left(\frac{M}{M_\oplus}\right)^{-0.15}
\ {\rm km\,s^{-1}}
\end{equation}

and

\begin{equation}
v_{\rm PL}(M)
=
4
\left(\frac{M}{M_\oplus}\right)^{-0.5}
\ {\rm km\,s^{-1}}.
\end{equation}

This confirms that lower-mass planets receive systematically larger ejection velocities, consistent with the mass-coupling prescriptions introduced above.

Spatially, the code places planets in a disk by drawing their radial distance $R$ from a gamma distribution with shape $2$ and scale length $R_d=3.5$~kpc, choosing the azimuthal angle $\phi$ uniformly from $0$ to $2\pi$, and drawing the vertical position $z$ from a Laplace distribution centered on the Galactic midplane with scale height $z_d=0.3$~kpc. These cylindrical coordinates are then converted into Cartesian positions $(x,y,z)$. 

Kinematically, the code gives the planets a warm, rotating disk velocity distribution rather than starting them with zero random motion. The local circular speed is calculated from the gravitational potential at each planet's actual position, including its height above the disk and its azimuthal location. Random radial and vertical velocities are then drawn from Gaussian distributions whose dispersions vary with radius, while the azimuthal velocity is centered below the circular speed using an asymmetric-drift correction and includes additional random motion with dispersion $0.7\sigma_R$. The resulting velocities are converted from cylindrical components $(v_R,v_\phi,v_z)$ into Cartesian components. Only after these initial positions and pre-kick velocities are assigned does the code add the ejection kick to create the kicked population; the matched control retains the original pre-kick velocities.

Since the FFPs are not started with exactly zero random motion, their radial and vertical velocities are drawn from Gaussian distributions, 

\begin{equation}
v_R\sim\mathcal{N}(0,\sigma_R)
\end{equation}

and 

\begin{equation}
v_z\sim\mathcal{N}(0,\sigma_z), 
\end{equation}

which gives the population realistic initial velocity dispersions. Their azimuthal velocity $v_\phi$ is centered around a mean rotation speed $\bar{v}_\phi$. In other words, $v_{\rm circ}$ provides the basic Galactic rotation, while $\sigma_R$, $\sigma_z$, and the asymmetric-drift correction give the FFP population a finite, dynamically warmer velocity distribution before the ejection kick is applied.

This implies that the final velocity is

\begin{equation}
\mathbf{v}_{\rm FFP}
=
\mathbf{v}_\star+\mathbf{v}_{\rm kick}.
\end{equation}

With the kick direction being isotropic:

\begin{equation}
\hat{\mathbf{n}}
=
(\sin\theta\cos\phi_k,
 \sin\theta\sin\phi_k,
 \cos\theta),
\end{equation}

with

\begin{equation}
\cos\theta\sim U(-1,1),
\qquad
\phi_k\sim U(0,2\pi).
\end{equation}

Therefore

\begin{equation}
\mathbf{v}_{\rm kick}
=
v_{\rm kick}\hat{\mathbf{n}}.
\end{equation}

The initial population is generated using a Jeans-like warm-disk prescription that provides nonzero radial and vertical random motions and an approximate rotational velocity. This initialization is designed to avoid the artificial collapse associated with the previous cold-disk setup, but it is not an exact equilibrium distribution function for the adopted triaxial Galactic potential.

\subsection{Gravitational potential of the Galaxy}
\label{sec:potential}
The Galactic potential is the sum of disk, bulge, and halo components \cite{McMillan_2016}, 
\begin{equation}
\Phi_{\rm tot}(x,y,z) = \Phi_{\rm disk}(R,z) + \Phi_{\rm bulge}(x,y,z) + \Phi_{\rm halo}(r), 
\end{equation}
where 
\begin{equation} R=\sqrt{x^2+y^2}, \qquad r=\sqrt{x^2+y^2+z^2}. \end{equation} 
The potential is static throughout the simulation. FFPs are treated as test particles and do not interact gravitationally with one another.

\subsection{Disk Potential}
\label{DiskPotential}

The disk potential used is a phenomenological softened potential of Miyamoto--Nagai-like form:
 
\begin{equation}
\Phi_{\rm disk}
=
-\frac{G M_{\rm disk}e^{-R/R_d}}
{\sqrt{R^2+(z_d+|z|)^2}}
\label{diskpotentialeq}
\end{equation}
 
where $M_{\rm disk}=6\times10^{10}\,M_\odot$.
 
\begin{table}[H] \centering \begin{tabular}{ll} \hline \textbf{Parameter} & \textbf{Value} \\ \hline $M_{\rm disk}$ & $6\times10^{10}M_\odot$ \\ $R_d$ & 3.5 kpc \\ $z_d$ & 0.3 kpc \\ \hline \end{tabular} \caption{Parameters of the implemented phenomenological disk potential.} \label{tab:disk_parameters} \end{table}
 
We note that Eq.~\eqref{diskpotentialeq} uses $|z|$ rather than $z^2$ in the vertical softening term, which departs from the standard Miyamoto--Nagai form (where the vertical dependence enters through $\sqrt{z^2+z_d^2}$ smoothly) and introduces a formal non-differentiability of the vertical force at $z=0$. A Miyamoto-Nagai-like disk with an added absolute-value term is not textbook-standard notation, and this is a deliberate departure adopted for numerical convenience rather than a literal Miyamoto-Nagai potential; this cusp is smoothed at the resolution scale $h$; particles crossing $z=0$ therefore experience a well-defined, continuous numerical force at all times, though a particle passing exactly through the midplane within a single finite-difference evaluation could in principle see a small numerical discontinuity smaller than $h$.
 
The key velocity scale is the local circular velocity.
 
\begin{equation}
v_{\rm circ}
=
\sqrt{
R\frac{\partial\Phi_{\rm tot}}{\partial R}
}
\end{equation}
 
Numerically,
 
\begin{equation}
\frac{\partial\Phi}{\partial R}
\simeq
\frac{\Phi(R+h)-\Phi(R-h)}{2h}.
\end{equation}
 
Thus,
 
\begin{equation}
v_{\rm circ}
=
\sqrt{
R
\frac{\Phi(R+h)-\Phi(R-h)}
{2h}
}.
\end{equation}
 
This is the $V_0$-type quantity that sets the underlying Galactic rotation before the FFP ejection kick. The code actually evaluates an \textbf{angle-aware} $v_{\rm circ}(x,y,z)$, because the triaxial bulge means that the circular-speed reference can vary with azimuth.

\subsection{Bulge Potential (Bissantz--Gerhard)}

The Galactic bulge is represented by a softened triaxial
phenomenological potential. We first rotate the Cartesian coordinates
into the principal-axis frame of the bulge,
\begin{align}
x' &= x\cos\theta_{\rm b}+y\sin\theta_{\rm b},\\
y' &= -x\sin\theta_{\rm b}+y\cos\theta_{\rm b},
\end{align}
where $\theta_{\rm b}$ is the orientation angle of the major axis in
the Galactic plane. The dimensionless ellipsoidal radius is then
\begin{equation}
s_{\rm b}
=
\left[
\left(\frac{x'}{X_{\rm b}}\right)^2
+
\left(\frac{y'}{Y_{\rm b}}\right)^2
+
\left(\frac{z}{Z_{\rm b}}\right)^2
\right]^{1/2}.
\end{equation}
Because $s_{\rm b}$ is dimensionless, we introduce the characteristic
bulge length
\begin{equation}
R_{\rm b}
=
\left(X_{\rm b}Y_{\rm b}Z_{\rm b}\right)^{1/3},
\end{equation}
and define the corresponding physical ellipsoidal radius as
\begin{equation}
r_{\rm b}=R_{\rm b}s_{\rm b}.
\end{equation}
The bulge potential is therefore written as
\begin{equation}
\Phi_{\rm b}(x,y,z)
=
-\frac{
4\pi G\rho_{\rm b,0}X_{\rm b}Y_{\rm b}Z_{\rm b}
}{
r_{\rm b}+a_{\rm b}
},
\label{eq:bulge_potential}
\end{equation}
where $\rho_{\rm b,0}$ is the bulge density scale and $a_{\rm b}$ is
the softening length. This form ensures that the denominator has units
of length and that $\Phi_{\rm b}$ has units of
$\mathrm{kpc}\,(\mathrm{km\,s^{-1}})^2$ when $G$ is expressed in
$\mathrm{kpc}\,(\mathrm{km\,s^{-1}})^2\,M_\odot^{-1}$.

\begin{table}[h]
\centering
\caption{Parameters used for the softened triaxial bulge potential.}
\label{tab:bulge_parameters}
\begin{tabular}{lcc}
\hline
Parameter & Value & Units \\
\hline
$\rho_{\rm b,0}$ & $10^{10}$ & $M_\odot\,\mathrm{kpc}^{-3}$ \\
$X_{\rm b}$ & $1.49$ & $\mathrm{kpc}$ \\
$Y_{\rm b}$ & $0.58$ & $\mathrm{kpc}$ \\
$Z_{\rm b}$ & $0.40$ & $\mathrm{kpc}$ \\
$a_{\rm b}$ & $0.50$ & $\mathrm{kpc}$ \\
$\theta_{\rm b}$ & $25$ & $\mathrm{deg}$ \\
\hline
\end{tabular}
\end{table}

The accelerations due to the bulge, disk, and halo are obtained by
centered finite differences of the total potential. In particular, the
dimensionless ellipsoidal coordinate used in the bulge model is first
converted to the physical length $r_{\rm b}=R_{\rm b}s_{\rm b}$ before
the softened denominator is evaluated.

\subsection{Halo Potential (NFW)}
\label{sec:Halo}

The dark-matter halo is represented by an NFW-like spherical potential,

\begin{equation}
\Phi_{\rm halo}
=
-GM_{\rm halo}
\frac{\ln\!\left(1+r_{\rm eff}/R_s\right)}
{r_{\rm eff}}
\end{equation}

where

\begin{equation}
r_{\rm eff}
=
\sqrt{r^2+R_{0,\rm halo}^2}.
\end{equation}

The small value of $R_{0,\rm halo}$ acts only as a numerical guard against the singularity at $r=0$. 

\begin{table}[H] 
\centering 
\begin{tabular}{ll} 
\hline \textbf{Parameter} & \textbf{Value} \\ \hline $M_{\rm halo}$ & $7.25\times10^{11}M_\odot$ \\ $R_s$ & 10 kpc \\ $R_{0,\rm halo}$ & $10^{-8}$ kpc \\ \hline 
\end{tabular} 
\caption{Parameters of the implemented halo potential.} 
\label{tab:halo_parameters} 
\end{table}

\subsection{Gravitational Acceleration} 
\label{sec:acceleration} 

The acceleration is obtained numerically from the gradient of the total potential using a centered finite-difference stencil. As shown, the equations of motion are

\begin{equation}
\mathbf{a}
=
-\nabla\Phi_{\rm tot}
\end{equation}

but we evaluate the derivatives numerically:

\begin{equation}
a_x
\simeq
-\frac{\Phi(x+h,y,z)-\Phi(x-h,y,z)}{2h},
\end{equation}

\begin{equation}
a_y
\simeq
-\frac{\Phi(x,y+h,z)-\Phi(x,y-h,z)}{2h},
\end{equation}

\begin{equation}
a_z
\simeq
-\frac{\Phi(x,y,z+h)-\Phi(x,y,z-h)}{2h},
\end{equation}

with $h=0.01$ kpc.

We designed a convergence test for the force calculation, repeating the integration with $h=0.005$, $0.01$, $0.02$, and $0.04$~kpc and comparing the resulting final-time $\sigma_R$ and $\sigma_z$ across the four runs (methodology and definitions in Appendix~\ref{app:integration}). The appendix accordingly reports the comparison.

\subsection{Galactic Escape Diagnostic} 
\label{sec:galactic_escape} 
The gravitational binding of each FFP is evaluated using its instantaneous specific orbital energy \citep{BinneyTremaine2008}, 
\begin{equation} 
E_i = \frac12 \left( v_{x,i}^2+v_{y,i}^2+v_{z,i}^2 \right) + \Phi_{\rm tot}(x_i,y_i,z_i). 
\label{eq:specific_energy} 
\end{equation} 

With the adopted convention $\Phi_{\rm tot}\rightarrow0$ at large radius, an object is formally unbound from the adopted Galactic potential when \begin{equation} 
E_i>0. 
\label{eq:unbound_condition} 
\end{equation} 

The corresponding local escape speed is 
\begin{equation}
v_{\rm esc}(\mathbf x) = \sqrt{-2\Phi_{\rm tot}(\mathbf x)}. \label{eq:escape_velocity} 
\end{equation} 

The Galactic escape fraction is therefore 
\begin{equation} 
f_{\rm Gal,esc}(t) = \frac{ N[E_i(t)>0] }{ N_{\rm tot} }. \label{eq:galactic_escape_fraction} 
\end{equation} 

This diagnostic concerns the total post-kick orbital energy. It is distinct from the simpler quantity $v_{\rm kick}>v_{\rm esc}$ because the FFP inherits the Galactic velocity of its parent stellar population and the kick is added vectorially \citep{Piffl2014,Monari2018,Deason2019}. A second diagnostic tracks departure from the thin-disk region. We define 
\begin{equation} 
f_{\rm disk,dep}(M,t) = P\left(|z|>2~{\rm kpc}\mid M,t\right), \label{eq:disk_departure_fraction}
\end{equation} 

where the threshold is 
\begin{equation}
Z_{\rm escape}=2~{\rm kpc}. 
\end{equation} 

This quantity does not represent gravitational escape from the Galaxy. It simply measures the fraction of particles that have moved beyond the adopted disk-departure height at a given stored snapshot. 

\subsection{Integration Scheme} 
\label{sec:integration} 

The fiducial simulation follows $10^6$ test particles for 
\begin{equation} T=10^8~{\rm yr}, 
\end{equation} 
using $10^4$ integration steps and therefore 
\begin{equation} 
\Delta t=10^4~{\rm yr}. 
\end{equation} 

We note that $10^8$~yr is short compared to the age of the Galactic disk ($\sim10^{10}$~yr); it is chosen as a computationally tractable timescale, following similar choices adopted in other sources (\citep{BhaskarPerets2025}). It does not by itself represent the full dynamical age over which a present-day FFP population would actually be observed, and we return to this point in the Conclusion \ref{sec:conclusion}.

The positions and velocities are advanced with a kick--drift--kick leapfrog scheme, 
\begin{align} \mathbf v_{n+1/2} &= \mathbf v_n+ \frac12\Delta t\,\mathbf a_n, \\ \mathbf x_{n+1} &= \mathbf x_n+ \Delta t\,\mathbf v_{n+1/2}, \\ \mathbf v_{n+1} &= \mathbf v_{n+1/2} + \frac12\Delta t\,\mathbf a_{n+1}. 
\end{align}

The physical time step is converted to the internal units of ${\rm kpc}/({\rm km\,s^{-1}})$ before the leapfrog update. The Galactic potential remains fixed throughout the integration, and the FFPs do not interact gravitationally with each other. The simulation stores the initial, midway, and final snapshots. The midway snapshot occurs at \begin{equation} t=5\times10^7~{\rm yr} \end{equation} for the fiducial $10^8$ yr integration.

\section{Diagnostics}
\label{sec:diagnostics}

\subsection{Propagation and Spatial Diagnostics} \label{sec:propagation_diagnostics}

The FFPs are propagated through the Galactic potential for $10^8$ years using a leapfrog kick-drift-kick integration scheme with a timestep of $10^4$ years. At each timestep, the gravitational accelerations are calculated from the gradient of the total disk, bulge, and halo potential using centered finite differences. As described in Sec.~\ref{sec:numerical_setup}, we track the three-dimensional positions and velocities of every FFP, allowing their spatial and kinematic evolution to be evaluated over time. Spatial diagnostics include the cylindrical galactocentric radius ($R=\sqrt{x^2+y^2}$), vertical displacement ($|z|$), and the fraction of objects reaching the adopted disk-departure threshold.

These diagnostics distinguish ordinary vertical or radial excursions within the Galactic potential from genuine gravitational escape and are compared between the kicked population and the matched no-kick control.

\subsection{Net-Displacement Diagnostic}
\label{sec:displacement_diagnostic}

To connect the injected ejection velocities to the resulting spatial distribution, we track the net displacement of each FFP from its birth location throughout the Galactic integration. For an FFP with position $\mathbf{r}_i(t)$, the displacement from its initial position is defined as
\begin{equation}
\Delta r_i(t) = \left|\mathbf{r}_i(t)-\mathbf{r}_i(0)\right|
= \left\{[x_i(t)-x_i(0)]^2+[y_i(t)-y_i(0)]^2+[z_i(t)-z_i(0)]^2\right\}^{1/2}.
\label{eq:net_displacement}
\end{equation}

The Net-Displacement Diagnostic is a way to measure how much an ejection kick changes a free-floating planet's location compared with where that same planet would have gone without the kick. To do this, the simulation follows two versions of each planet: a kicked planet, which receives an initial ejection velocity, and a no-kick control planet, which starts with the same position, mass, and original velocity but receives no additional kick. Both are then moved through the same Galactic gravitational potential. By comparing their positions at the same time, the simulation calculates the separation

\begin{equation}
\Delta r_{\rm kick}(t)
=
\left|
\mathbf{r}_{\rm kicked}(t)
-
\mathbf{r}_{\rm control}(t)
\right|.
\end{equation}

This removes the motion that both planets share because of their normal Galactic orbit, allowing the diagnostic to focus specifically on the spatial effect of the ejection kick. The simulation measures this separation at $1$, $10$, $30$, and $100$~Myr and summarizes the results for different planet-mass bins using the median and 90th-percentile separations.

\subsection{Mass-Resolved Diagnostics} 
\label{sec:mass_resolved_diagnostics} 

The Mass-Resolved Diagnostics are designed to determine whether the FFPs' mass is associated with different kick strengths and different spatial outcomes. Following the setup of Sec.~\ref{sec:numerical_setup}, we divide the planets into five Earth-mass bins,
 
\begin{equation}
1\!-\!3\,M_\oplus, 3\!-\!10\,M_\oplus, 10\!-\!30\,M_\oplus, 30\!-\!100\,M_\oplus, and 100\!-\!317\,M_\oplus;
\end{equation}
 
the final bin is inclusive of the $317\,M_\oplus$ upper limit.
 
These same bins are used consistently for the kick-speed statistics, disk-departure fractions, and spatial $(R,z)$ maps.
 
We then map the resulting spatial distribution in galactocentric radius $R$ and height $z$, showing the kicked population alongside the matched no-kick control for each mass bin.
 
These maps are intended to trace $P(R,z\mid M,t)$, allowing us to see whether particular mass ranges become preferentially displaced from the Galactic disk as a result of the kicks.
 
Finally, the we calculate the fraction of each mass bin that reaches $|z|>2$ kpc, providing a mass-dependent measure of disk departure rather than true gravitational escape.
 
\section{Numerical Results}
\label{sec:results}
 
\subsection{Table Quantities}
\label{sec:results_spatial}
 
\begin{table}[H]
\centering
\renewcommand{\arraystretch}{1.4}
\begin{tabular}{|p{0.28\textwidth}|p{0.58\textwidth}|}
\hline
N & Number of FFPs in the sample or mass bin \\
\hline
$\sigma_R$ & Radial velocity dispersion (random radial motion) \\
\hline
$\sigma_\phi$ & Azimuthal velocity dispersion (spread in orbital speed) \\
\hline
$\sigma_z$ & Vertical velocity dispersion (random motion above/below the disk plane) \\
\hline
 
\end{tabular}
\caption{Summary of diagnostic quantities used in the simulation analysis. As implemented numerically (see the note following Table~\ref{tab:final_velocity_comparison}), $\sigma_\phi$ is computed as the dispersion of $v_\phi-v_{\rm circ}(x,y,z)$ -- the local circular velocity, not the population's local mean streaming velocity $\overline{v_\phi}(R)$ -- and so retains a contribution from the asymmetric-drift lag of the warm-disk initial condition rather than being a pure random-motion dispersion.}
\label{tab:diagnostic_quantities}
\end{table}

The quantities listed in Table~\ref{tab:diagnostic_quantities} provide complementary measures of the FFP population's dynamical state. The velocity dispersions $\sigma_R$, $\sigma_\phi$, and $\sigma_z$ are calculated from the simulated particle velocities and quantify radial, azimuthal, and vertical random motion, respectively. The spatial diagnostics track the corresponding redistribution in the Galactic potential, while the disk-departure and escape diagnostics distinguish vertical excursions from objects that become formally unbound. Together, these quantities allow the injected ejection velocity to be followed from the initial kick through its subsequent dynamical evolution.

\subsection{Disk Departure and Time-Evolution Diagnostics}
\label{sec:results_kick_escape_time} 

\begin{table}[H]
\centering
\caption{Disk-departure fraction by FFP mass and ejection channel, defined as the fraction of objects satisfying $|z|>2~\mathrm{kpc}$ at $t=10^8$ yr. The final row gives the overall fraction for each channel, independent of mass.}
\label{tab:disk_departure_mass_final}

\begin{tabular}{lcccc}
\hline
\textbf{Mass bin [$M_\oplus$]} & \textbf{CB -- null} & \textbf{CB -- coupled} & \textbf{PL -- null} & \textbf{PL -- coupled} \\
\hline
1--3       & 0.033 & 0.032 & 0.028 & 0.028 \\
3--10      & 0.032 & 0.031 & 0.029 & 0.029 \\
10--30     & 0.034 & 0.032 & 0.029 & 0.029 \\
30--100    & 0.029 & 0.030 & 0.028 & 0.028 \\
100--317   & 0.000 & 0.000 & 0.029 & 0.028 \\
\hline
\textbf{Overall} & \textbf{0.033} & \textbf{0.032} & \textbf{0.028} & \textbf{0.028} \\
\hline
\end{tabular}

\end{table}

Table~\ref{tab:disk_departure_mass_final} reports the disk-departure fraction $f_{\rm disk,dep}$ (Eq.~\ref{eq:disk_departure_fraction}) by mass bin and channel at $t=10^8$ yr. Departure fractions stay at the few-percent level throughout, with overall values of $0.033$ (null) and $0.032$ (coupled) for CB, and $0.028$ for PL in both models -- indicating the mass-dependent kick barely shifts the global disk-departure rate. The zero value in the $100$--$317\,M_\oplus$ CB bin reflects its small sample ($N=56$; Table~\ref{tab:mass_binned_kicks}) rather than a genuinely vanishing departure probability.

\begin{table}[H]
\centering
\caption{Time evolution of the principal kinematic and spatial diagnostics for the null and mass-coupled models.}
\label{tab:time_evolution}
\begin{tabular}{lccc}
\hline
\textbf{Quantity} & \textbf{Initial} & \textbf{Midway} & \textbf{Final} \\
\hline
$\sigma_R$ --- null / coupled & 54.12 / 54.02 & 83.39 / 83.25 & 87.67 / 87.57 \\
$\sigma_z$ --- null / coupled & 32.32 / 32.14 & 78.52 / 78.48 & 82.05 / 82.01 \\
$\langle|z|\rangle \text{ (mean absolute height)}$ --- null / coupled [kpc] & 0.300 / 0.300 & 0.420 / 0.416 & 0.494 / 0.490 \\
$f_{\rm disk,dep}$ --- null / coupled & 0.001 / 0.001 & 0.011 / 0.011 & 0.031 / 0.030 \\
\hline
\end{tabular}
\end{table}

Table~\ref{tab:time_evolution} tracks $\sigma_R$, $\sigma_z$, $\langle|z|\rangle$, and $f_{\rm disk,dep}$ at the initial, midway, and final snapshots. Both models broaden substantially: $\sigma_R$ rises from $\sim54$ to $\sim88~{\rm km\,s^{-1}}$ and $\sigma_z$ from $\sim32$ to $\sim82~{\rm km\,s^{-1}}$, while $\langle|z|\rangle$ grows from $0.30$ to $\sim0.49$~kpc. Because the null and coupled values agree closely at every snapshot, this bulk evolution is shared by both models. The similarity indicates that the global kinematic evolution is dominated by the subsequent motion of the initially warm population in the Galactic potential, rather than by differences between the two ejection prescriptions. The specific contribution of the initial kick is isolated more directly through the matched no-kick control comparison in Sec.~\ref{sec:nokick_control_results}.

\subsection{Velocity Evolution} 
\label{sec:results_velocity} 

\begin{table}[H]
\centering
\caption{Final velocity dispersions, estimated vertical scale heights, and anisotropy ratios for the null and mass-coupled models. Velocity dispersions are given in km\,s$^{-1}$.}
\label{tab:final_velocity_comparison}
\begin{tabular}{llccccc}
\hline
\textbf{Model} & \textbf{Ch} & \textbf{$\sigma_R$} & \textbf{$\sigma_\phi$} & \textbf{$\sigma_z$} & \textbf{$\langle|z|\rangle $ [kpc]} & \textbf{$\sigma_R/\sigma_z$} \\
\hline
null     & CB  & 88.18 & 92.89 & 82.86 & 0.512 & 1.06 \\
null     & PL  & 87.15 & 92.44 & 81.24 & 0.475 & 1.07 \\
null     & ALL & 87.67 & 92.66 & 82.05 & 0.494 & 1.07 \\
coupled  & CB  & 88.02 & 92.83 & 82.89 & 0.505 & 1.06 \\
coupled  & PL  & 87.11 & 92.37 & 81.11 & 0.474 & 1.07 \\
coupled  & ALL & 87.57 & 92.60 & 82.01 & 0.490 & 1.07 \\
\hline
\end{tabular}
\end{table}

Table~\ref{tab:final_velocity_comparison} gives the final-snapshot $\sigma_R$, $\sigma_\phi$, $\sigma_z$, $\langle|z|\rangle$, and anisotropy ratio $\sigma_R/\sigma_z$ per channel and model. Null and coupled values agree closely, with combined $\sigma_R\simeq87.7$ versus $87.6~{\rm km\,s^{-1}}$ and $\sigma_z\simeq82.1$ versus $82.0~{\rm km\,s^{-1}}$. The ratio $\sigma_R/\sigma_z\simeq1.07$ is also nearly identical in the two models. These results show that the mass-coupled prescription leaves the global final velocity moments essentially unchanged relative to the null model. They do not, however, imply that the population remains close to its initial kinematic state, since both models undergo substantial evolution from the initially warm-disk configuration.

\subsection{Mass Dependence Injected at Ejection}
\label{sec:results_mass_injection}

\begin{table}[H]
\centering
\caption{Injected kick speed vs.\ FFP mass. Values are sampled medians in km\,s$^{-1}$; $N$ is the number of particles in each mass bin.}
\label{tab:mass_binned_kicks}
\begin{tabular}{lrrrrrr}
\hline
\textbf{Mass bin [$M_\oplus$]} & \textbf{$N$ (CB)} & \textbf{CB -- null} & \textbf{CB -- coupled} & \textbf{$N$ (PL)} & \textbf{PL -- null} & \textbf{PL -- coupled} \\
\hline
1--3 & 383493 & 15.381 & 14.320 & 170602 & 3.998 & 3.121 \\
3--10 & 106085 & 15.366 & 12.256 & 132601 & 4.001 & 1.771 \\
10--30 & 9579 & 15.440 & 10.343 & 85428 & 4.002 & 1.046 \\
30--100 & 787 & 15.346 & 8.753 & 66615 & 3.993 & 0.789 \\
100--317 & 56 & 15.470 & 7.326 & 44754 & 4.014 & 0.708 \\
\hline
\end{tabular}
\end{table}

Table~\ref{tab:mass_binned_kicks} reports the sampled median kick speed per mass bin and channel---the direct numerical check of the mass-coupling prescriptions of Sec.~\ref{sec:kick_initialization}. The null-model medians remain approximately constant across the mass bins, at $\sim15.4~{\rm km\,s^{-1}}$ for CB and $\sim4.0~{\rm km\,s^{-1}}$ for PL. In the coupled model, the median kick speed decreases systematically with mass, from $14.32$ to $7.33~{\rm km\,s^{-1}}$ for CB and from $3.12$ to $0.71~{\rm km\,s^{-1}}$ for PL. This establishes that the intended mass--velocity correlation is present in the sampled initial kicks. The subsequent spatial and kinematic signatures must nevertheless be interpreted together with the evolution of the matched no-kick control population, since the Galactic potential also redistributes the initially warm population during the integration.

The sampled coupled-model PL medians are broadly consistent with the adopted scaling $$v_{\rm PL,base}(M)=4(M/M_\oplus)^{-1/2}~{\rm km\,s^{-1}}$$ in the three lowest mass bins. Using the representative masses implied by the adopted $dN/dM\propto M^{-1.3}$ distribution gives expected median speeds of approximately $3.1$, $1.8$, $1.0$, $0.56$, and $0.31~{\rm km\,s^{-1}}$ across the five mass bins. The corresponding sampled values in Table~\ref{tab:mass_binned_kicks} are $3.121$, $1.771$, $1.046$, $0.789$, and $0.708~{\rm km\,s^{-1}}$: the first three bins agree at the few-percent level, but the two highest-mass bins depart from the pure power-law expectation by roughly $40\%$ and a factor of $\sim2.3$, respectively. This flattening at high mass indicates that the sampled PL kick distribution does not track the ideal $M^{-1/2}$ scaling over the full mass range implemented, and is worth revisiting in the sampling code.

\subsection{Propagation Distance and Mass-Dependent Spatial Redistribution}
\label{sec:results_displacement}

\begin{figure}[H]
    \centering
    \includegraphics[width=0.7\linewidth]{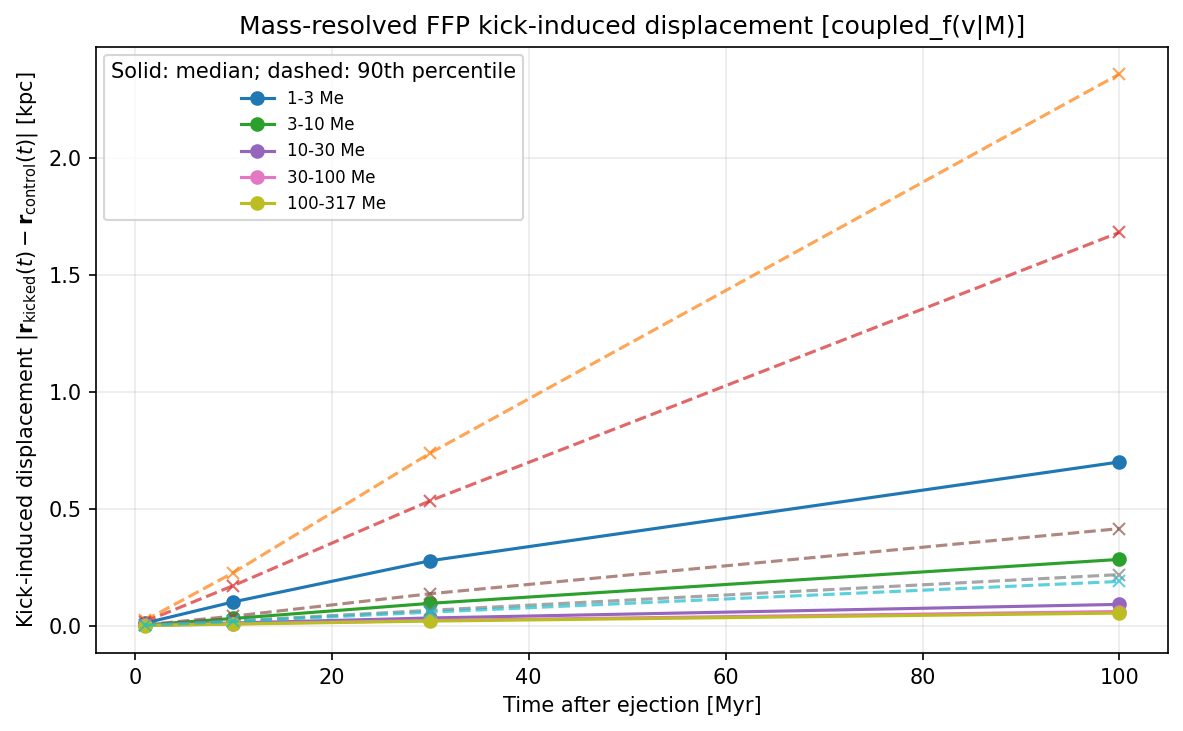}
    \caption{Mass-coupled model: median (solid) and 90th-percentile (dashed) net displacement $\Delta r(t)$ as a function of post-ejection propagation time, for each of the five FFP mass bins.}
    \label{fig:displacement_coupled}
\end{figure}

\begin{figure}[H]
    \centering
    \includegraphics[width=0.7\linewidth]{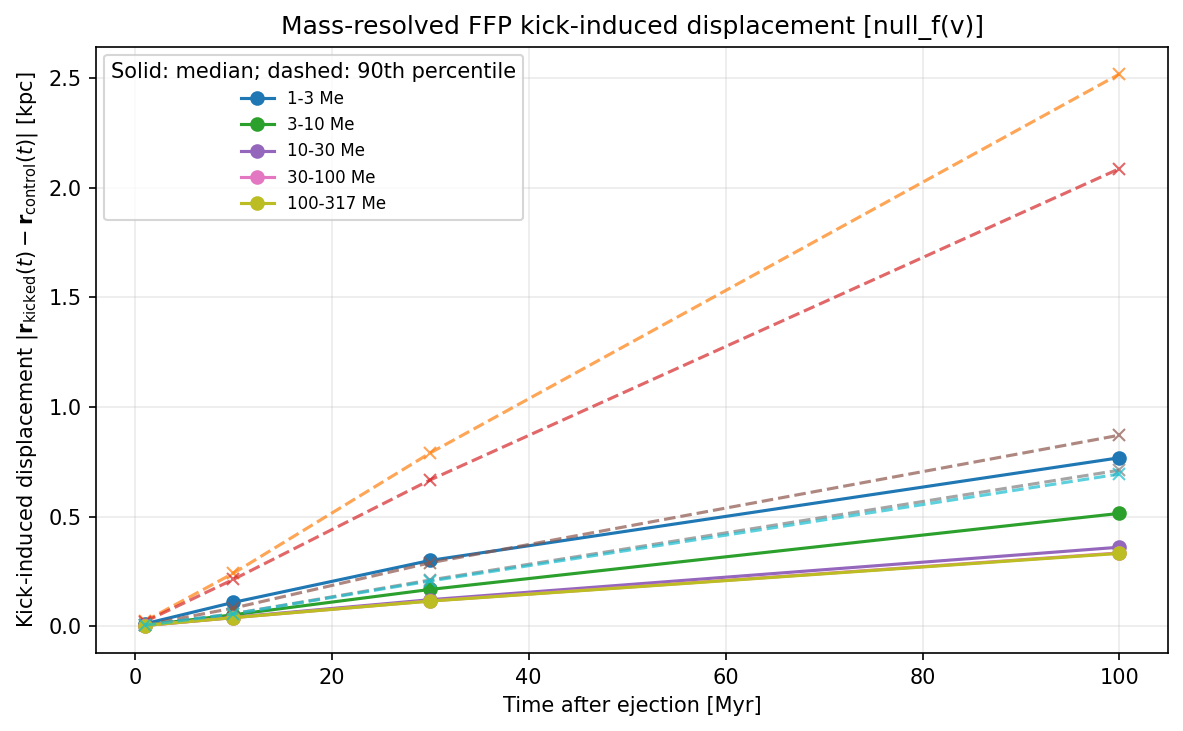}
    \caption{Null model: median (solid) and 90th-percentile (dashed) net displacement $\Delta r(t)$ as a function of post-ejection propagation time, for each of the five FFP mass bins.}
    \label{fig:displacement_null}
\end{figure}

To connect the injected ejection velocities (Section~\ref{sec:results_mass_injection}) to the resulting spatial distribution, we examine the net-displacement diagnostic $\Delta r_i(t)$ introduced in Section~\ref{sec:displacement_diagnostic} (Eq.~\ref{eq:net_displacement}). Figures~\ref{fig:displacement_null} and \ref{fig:displacement_coupled} show the median and 90th-percentile displacement, $\Delta r_{50}(M,t)$ and $\Delta r_{90}(M,t)$, as a function of post-ejection propagation time for each of the five FFP mass bins, separately for the null and mass-coupled kick prescriptions.

The coupled ejection model produces a stronger mass-dependent hierarchy in the matched kicked-versus-control displacement than the null model. However, the null model also exhibits a mass dependence in the final displacement despite its approximately mass-independent kick prescription. This indicates that the matched displacement reflects both the injected kick distribution and the subsequent dynamical evolution of the population, while the stronger trend in the coupled model is consistent with the imposed mass--kick correlation.

\begin{table}[H]
\centering
\caption{Combined-population net displacement as a function of post-ejection propagation time, for the null and mass-coupled models.}
\label{tab:displacement_overall}

\begin{tabular}{lcccccc}
\hline
\textbf{$t$ [Myr]} &
\multicolumn{3}{c}{\textbf{Null}} &
\multicolumn{3}{c}{\textbf{Coupled}} \\
&
$\Delta r_{50}$ &
$\Delta r_{90}$ &
$R_{\rm GC}^{\rm med}$ &
$\Delta r_{50}$ &
$\Delta r_{90}$ &
$R_{\rm GC}^{\rm med}$ \\
\hline
1   & 0.0058 & 0.0220 & 5.896 & 0.0047 & 0.0198 & 5.896 \\
10  & 0.0594 & 0.2221 & 5.928 & 0.0452 & 0.2002 & 5.928 \\
30  & 0.1889 & 0.7024 & 6.235 & 0.1340 & 0.6330 & 6.234 \\
100 & 0.5563 & 2.2062 & 5.941 & 0.3720 & 1.9938 & 5.940 \\
\hline
\end{tabular}

\end{table}

Table~\ref{tab:displacement_overall} lists the combined-population $\Delta r_{50}$, $\Delta r_{90}$, and $R_{\rm GC}^{\rm med}$ (Eq.~\ref{eq:net_displacement}) at $t=1$--$100$~Myr. The matched kicked-versus-control displacement grows with propagation time in both models, from median values of $0.0058$ and $0.0047$~kpc at $1$~Myr to $0.5563$ and $0.3720$~kpc at $100$~Myr for the null and coupled models, respectively.

\begin{table}[H]
\centering
\caption{Mass-resolved net displacement at the final propagation time, $t=100$~Myr, compared against the injected combined-channel (CB+PL) median kick speed for the same mass bin (cf.\ Table~\ref{tab:mass_binned_kicks}). $N$ is the combined CB+PL particle count in each mass bin. The final column gives the ratio of the coupled-model median kick in the lowest-mass bin (1--3~$M_\oplus$) to that in the highest-mass bin (100--317~$M_\oplus$).}
\label{tab:displacement_mass_resolved}

\begin{tabular}{lrrrrrr}
\hline
\textbf{Mass bin [$M_\oplus$]} &
\textbf{$N$} &
\multicolumn{2}{c}{\textbf{Null}} &
\multicolumn{2}{c}{\textbf{Coupled}} &
\textbf{Range ratio} \\
\cline{3-6}
&
&
$v_{\rm kick}^{\rm med}$ [km/s] &
$\Delta r_{50}$ [kpc] &
$v_{\rm kick}^{\rm med}$ [km/s] &
$\Delta r_{50}$ [kpc] &
(coupled) \\
\hline
1--3     & 554095 & 11.519 & 0.7687 & 10.716 & 0.7001 & \multirow{5}{*}{$15.1\times$} \\
3--10    & 238686 & 5.135  & 0.5153 & 3.035  & 0.2835 & \\
10--30   & 95007  & 4.140  & 0.3611 & 1.173  & 0.0911 & \\
30--100  & 67402  & 4.007  & 0.3349 & 0.800  & 0.0605 & \\
100--317 & 44810  & 4.015  & 0.3329 & 0.708  & 0.0539 & \\
\hline
\end{tabular}

\end{table}

Table~\ref{tab:displacement_mass_resolved} shows that the mass dependence of the injected kick is reflected in the matched kicked-versus-control displacement at the resolved-bin level. In the coupled model, the median injected kick speed decreases from $10.716$ to $0.708~{\rm km\,s^{-1}}$ across the five mass bins, corresponding to a factor of $\sim15.1$ range. Over the same mass range, the final median displacement decreases from $0.7001$ to $0.0539$~kpc. The null model also shows a decrease in median displacement, from $0.7687$ to $0.3329$~kpc, despite its approximately mass-independent kick prescription. Therefore, the final matched displacement retains a clear mass dependence in both models, but the trend is substantially stronger in the mass-coupled case. The null-model result demonstrates that mass-dependent spatial redistribution can arise from the sampled initial conditions and subsequent orbital evolution even when no explicit mass dependence is imposed on the kick speed.

\subsection{Mass-resolved spatial distribution}
\label{sec:MassResolvedSpatialResults}

The mass-resolved spatial analysis directly examines 
\begin{equation} 
P_{\rm FFP}(R,z\mid M,t) 
\end{equation} 
through two-dimensional kernel-density estimates. For each of the five mass bins, the kicked population is shown alongside the matched no-kick control at the initial, midway, and final snapshots. 

The spatial KDEs therefore provide a direct visual test of whether the injected mass--velocity correlation produces greater radial or vertical redistribution for lower-mass FFPs. The control population is the no-kick run, and the kicked population is the kicked run.

\begin{figure}[H]
    \centering
    \includegraphics[width=1.0\linewidth]{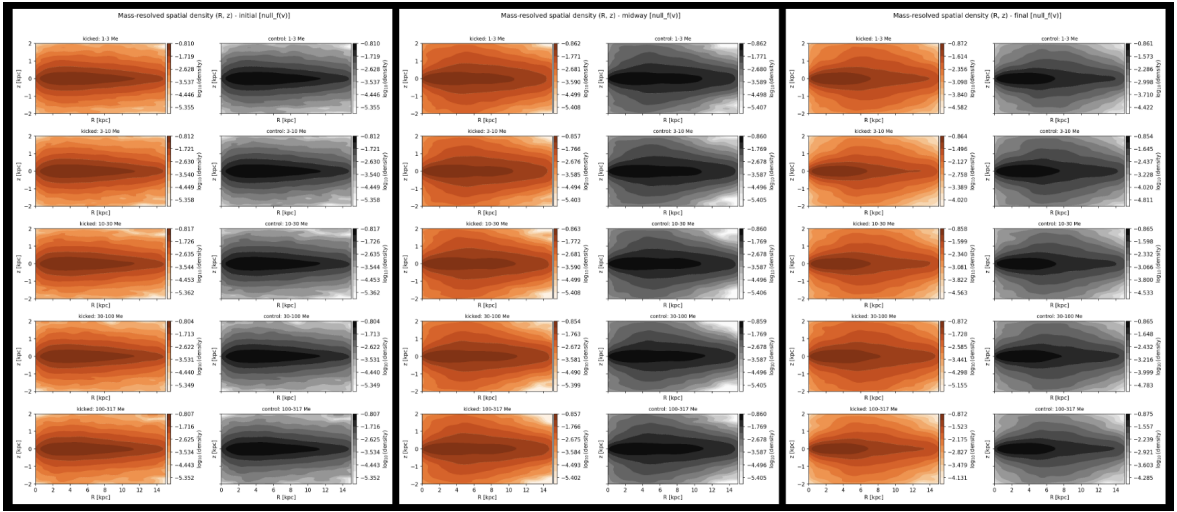}
    \caption{Null model: mass-resolved $(R,z)$ kernel-density estimate, kicked population vs.\ matched no-kick control, at the initial, midway, and final snapshots (panels left to right), one row per mass bin. In the null model, all masses within a given channel are drawn from the same kick-speed prescription.}
    \label{fig:massresolved_null}
\end{figure}

The null-model mass-resolved KDE provides the baseline spatial evolution against which the mass-coupled model can be compared. Because the kick prescription is independent of mass in this model, differences between mass bins are not caused by an imposed mass dependence in the ejection velocity. Instead, they can arise from the sampled initial conditions, the mass distribution across the bins, and the subsequent orbital evolution in the Galactic potential. The comparison with the matched no-kick population therefore establishes how much radial and vertical redistribution occurs without a mass-dependent kick prescription.

\begin{figure}[H]
    \centering
    \includegraphics[width=1.0\linewidth]{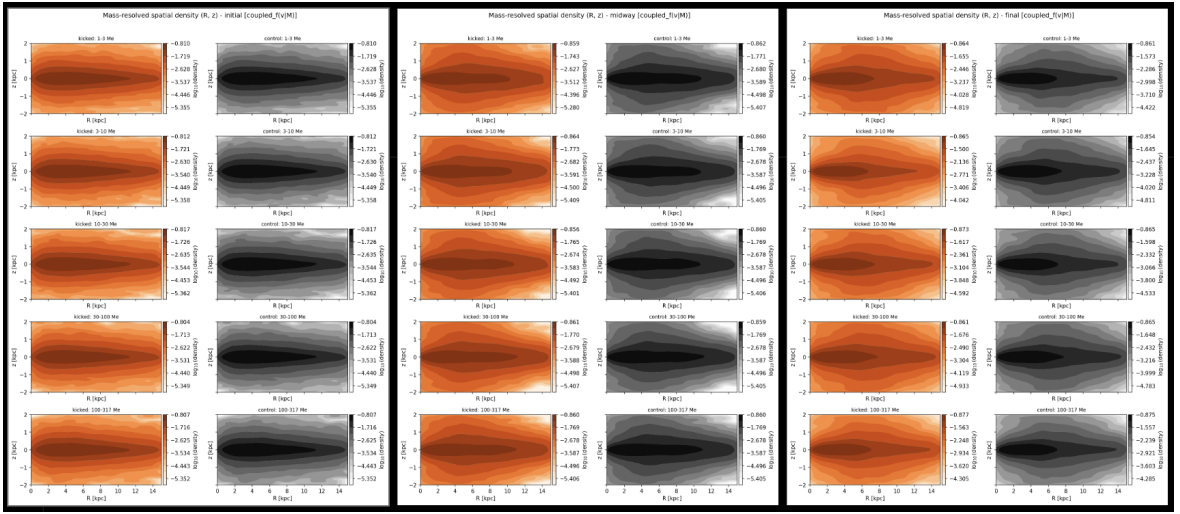}
    \caption{Mass-coupled model: mass-resolved $(R,z)$ kernel-density estimate, kicked population vs.\ matched no-kick control, at the initial, midway, and final snapshots (panels left to right), one row per mass bin. In the coupled model, lower-mass objects are assigned systematically larger characteristic kick scales.}
    \label{fig:massresolved_coupled}
\end{figure}

The coupled-model KDEs test whether the injected mass--velocity relation produces a corresponding mass dependence in the Galactic spatial distribution. The kick prescriptions assign systematically larger characteristic velocities to lower-mass FFPs, so the lowest-mass bins provide the strongest test for enhanced radial and vertical redistribution. The relevant comparison is therefore not simply the absolute width of each distribution, but the difference between the kicked population and its matched no-kick control at the same mass and time. This comparison helps separate the spatial effect of the injected velocity correlation from the background evolution of the initially warm population.

\subsection{Phase-Space Contours}
\label{sec:results_phase_space}

For each stored snapshot the code produces four two-dimensional KDE projections: 
\begin{enumerate}
\item 
The spatial distribution $(R,z)$; \item The radial phase-space distribution $(R,v_R)$; \item The vertical-heating distribution $(R,|v_z|)$; \item The velocity-space projection $(v_R,v_z)$. 
\end{enumerate} 

Each projection is shown separately for the CB and PL channels. When more than $50\,000$ particles are available in a channel, the KDE is evaluated using a random subset of $50\,000$ particles for computational efficiency. 

The density is evaluated on a $120\times120$ grid using a Gaussian kernel-density estimator and displayed through logarithmic density contours. 

\begin{figure}[H]
    \centering
    \includegraphics[width=1.0\linewidth]{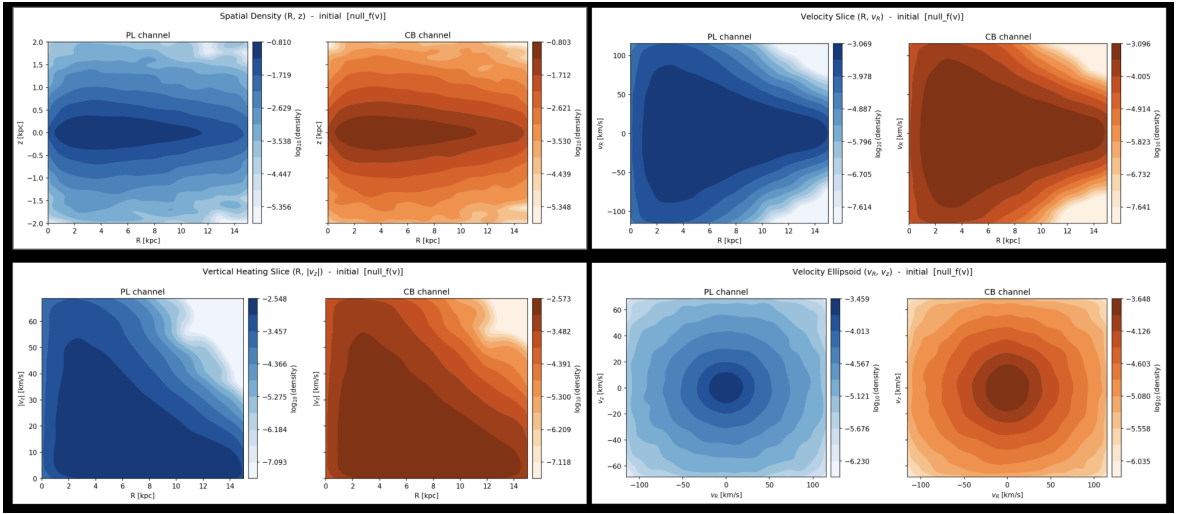}
    \caption{Null model, initial snapshot. Panels show, from left to right (repeated for the CB and PL channels, top and bottom): $(R,z)$, $(R,v_R)$, $(R,|v_z|)$, and $(v_R,v_z)$.}
    \label{fig:phasespace_null_initial}
\end{figure}

At the initial snapshot, the phase-space distributions establish the kinematic state immediately after initialization and ejection. The $(R,z)$ projection traces the spatial distribution of the population, while $(R,v_R)$ and $(R,|v_z|)$ quantify radial and vertical motions as a function of Galactocentric radius. The $(v_R,v_z)$ projection provides a direct view of the random velocity structure. These distributions form the baseline from which subsequent broadening and redistribution can be assessed during the $10^8$ yr integration. Because the initial conditions are generated using a Jeans-like warm-disk prescription rather than an exact equilibrium distribution function for the adopted triaxial potential, some subsequent evolution is expected even in the absence of ejection kicks.

\begin{figure}[H]
    \centering
    \includegraphics[width=1.0\linewidth]{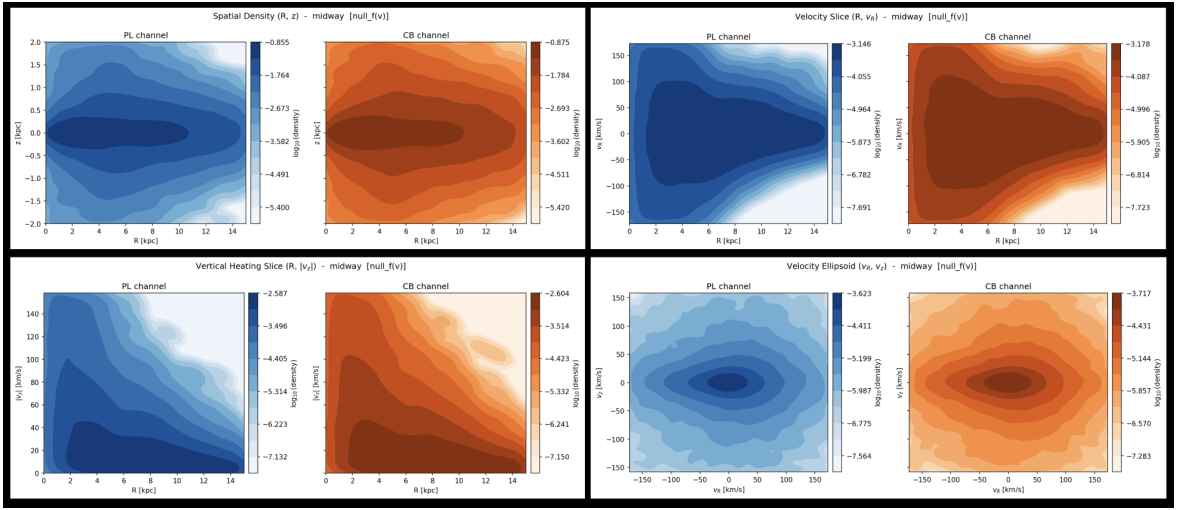}
    \caption{Null model, midway snapshot. Panels show, from left to right (repeated for the CB and PL channels, top and bottom): $(R,z)$, $(R,v_R)$, $(R,|v_z|)$, and $(v_R,v_z)$.}
    \label{fig:phasespace_null_midway}
\end{figure}

By the midpoint at $t=5\times10^7$ yr, the phase-space distributions have broadened as the initially warm population evolves through the Galactic potential. The changes in $(R,v_R)$ and $(R,|v_z|)$ trace the development of larger radial and vertical velocity excursions, while the $(R,z)$ distribution records the corresponding spatial redistribution. The midway snapshot therefore reflects a combination of the initial ejection perturbation and the ongoing dynamical evolution of the warm initial population. It provides an intermediate reference between the injected state and the accumulated evolution seen at the final time.

\begin{figure}[H]
    \centering
    \includegraphics[width=1.0\linewidth]{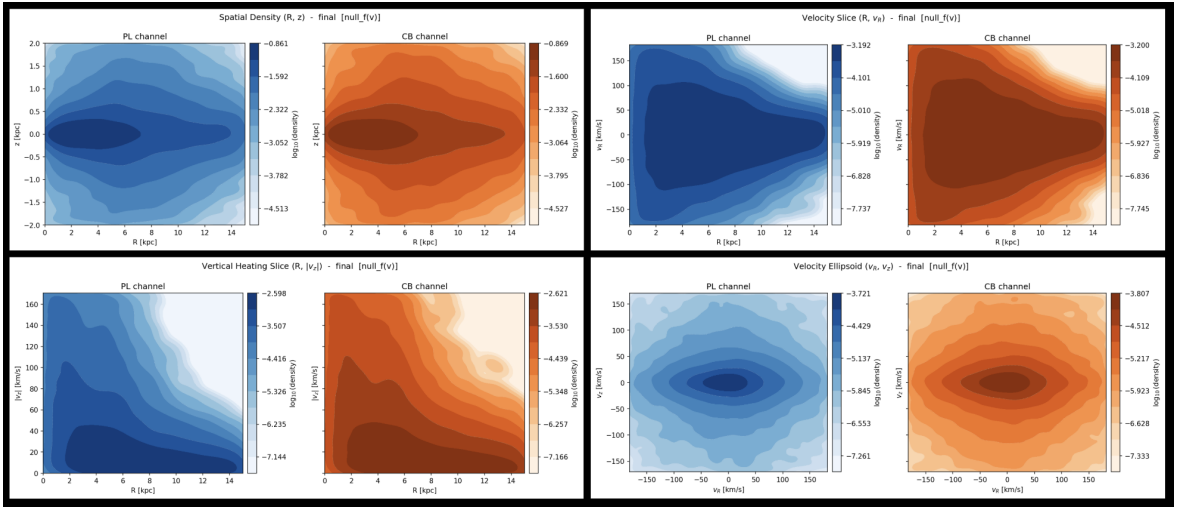}
    \caption{Null model, final snapshot. Panels show, from left to right (repeated for the CB and PL channels, top and bottom): $(R,z)$, $(R,v_R)$, $(R,|v_z|)$, and $(v_R,v_z)$.}
    \label{fig:phasespace_null_final}
\end{figure}

At $t=10^8$ yr, the phase-space projections show the accumulated response of the FFP population to both the initial ejection kick and subsequent motion in the Galactic potential. The broadening of the velocity-space distributions is consistent with the increase in the measured velocity dispersions, while the $(R,z)$ projection shows the associated redistribution away from the initial disk configuration. The matched no-kick control demonstrates that a substantial portion of this broadening and redistribution occurs even without ejection kicks. The final distributions should therefore be interpreted together with the control comparison and the energy-based escape diagnostic, since a broadened phase-space distribution does not by itself imply that the FFPs have escaped the Galaxy.

\subsection{Kicked-versus-Control Comparison}
\label{sec:nokick_control_results}
 
\begin{table}[H]
\centering
\caption{Final kinematic comparison between the kicked FFP populations and their matched no-kick controls at $t=10^8$ yr. The control populations have the same birth positions, masses, and pre-kick warm-disk velocities as the corresponding kicked populations, but no ejection kick. The columns $\sigma_R$ and $\sigma_z$ give the radial and vertical velocity dispersions, while $\Delta\sigma_R$ and $\Delta\sigma_z$ quantify the additional dispersion attributable to the kick in quadrature, defined as $\sqrt{\max(0,\sigma_{\rm kick}^2-\sigma_{\rm control}^2)}$.}
\label{tab:control_comparison}
 
\begin{tabular}{llcccccc}
\hline
\textbf{Model} & \textbf{Channel} &
$\boldsymbol{\sigma_R^{\rm kick}}$ &
$\boldsymbol{\sigma_R^{\rm ctrl}}$ &
$\boldsymbol{\Delta\sigma_R}$ &
$\boldsymbol{\sigma_z^{\rm kick}}$ &
$\boldsymbol{\sigma_z^{\rm ctrl}}$ &
$\boldsymbol{\Delta\sigma_z}$ \\
\hline
Null     & CB  & 88.18 & 87.17 & 13.33 & 82.86 & 82.47 & 7.98 \\
Null     & PL  & 87.15 & 87.19 & 0.00  & 81.24 & 81.29 & 0.00 \\
Null     & ALL & 87.67 & 87.18 & 9.23  & 82.05 & 81.88 & 5.29 \\
\hline
Coupled  & CB  & 88.02 & 87.17 & 12.26 & 82.89 & 82.47 & 8.36 \\
Coupled  & PL  & 87.11 & 87.19 & 0.00  & 81.11 & 81.29 & 0.00 \\
Coupled  & ALL & 87.57 & 87.18 & 8.26  & 82.01 & 81.88 & 4.54 \\
\hline
\end{tabular}
\end{table}
 
Table~\ref{tab:control_comparison} compares final $\sigma_R$ and $\sigma_z$ between each kicked population and its matched no-kick control (Appendix~\ref{app:control}), with $\Delta\sigma_R$ and $\Delta\sigma_z$ representing the kick-attributable quadrature excess. The null-model control alone reaches $\sigma_R=87.18$ and $\sigma_z=81.88~{\rm km\,s^{-1}}$, nearly matching the full kicked population with $\sigma_R=87.67$ and $\sigma_z=82.05~{\rm km\,s^{-1}}$. The corresponding excesses are only $9.23$ and $5.29~{\rm km\,s^{-1}}$ for the combined population. The coupled model gives similarly modest excesses of $8.26$ and $4.54~{\rm km\,s^{-1}}$. The residual heating is channel-dependent and is concentrated primarily in the CB population, where $\Delta\sigma_R\simeq12$--$13$ and $\Delta\sigma_z\simeq8~{\rm km\,s^{-1}}$, while the PL population shows zero excess at the quoted precision. Thus, most of the raw dispersion growth in Table~\ref{tab:time_evolution} is associated with the background evolution of the initially warm population rather than with the ejection kick alone. The matched control is therefore essential for identifying the comparatively modest additional kinematic contribution of the kick.

\section{Phase-Space Distribution Function and Connection to Microlensing}
\label{sec:Microlensing}

A central ingredient for predicting microlensing event rates is the phase-space distribution function (DF) of free-floating planets (FFPs),

\begin{equation}
f(\mathbf{x},\mathbf{v},t)=\frac{\mathrm{d}^{6}N}{\mathrm{d}^{3}\mathbf{x}\,\mathrm{d}^{3}\mathbf{v}},
\end{equation}

which encodes the number density of FFPs in position $\mathbf{x}$ and velocity $\mathbf{v}$. In equilibrium, $f$ is conserved along orbits in the Galactic potential according to the collision-free Boltzmann equation. The task is therefore to specify physically motivated initial conditions at the time of ejection and evolve them forward under the Galactic potential. 

In this work, we obtain this distribution function numerically by evolving large ensembles of test particles to a fixed final time, rather than solving the collisionless Boltzmann equation analytically; we note this final-time snapshot is not dynamically settled into equilibrium in the strict sense (Sec.~\ref{sec:nokick_control_results}).

\subsection{Symmetries and Approximations}
\label{sec:symmetries}

For the schematic microlensing calculation only, we adopt an axisymmetrized representation of the simulated phase-space distribution:

\begin{equation}
f(\mathbf{x},\mathbf{v}) = f(R,z;v_R,v_\phi,v_z),
\end{equation}

in cylindrical coordinates $(R,\phi,z)$, with $\phi$ suppressed due to axisymmetry. The velocity ellipsoid is not isotropic: stars and FFPs in the Galactic disk typically satisfy

\begin{equation}
\sigma_R > \sigma_\phi > \sigma_z,
\end{equation}

where $(\sigma_R,\sigma_\phi,\sigma_z)$ are the velocity dispersions in the respective components. This ordering reflects epicyclic motion, asymmetric drift, and vertical confinement of the disk. We therefore construct FFP DFs that inherit the same qualitative anisotropy as the stellar population. This approximation is used only to obtain a tractable shape-level microlensing calculation; the underlying Galactic dynamical integration includes the rotated triaxial bulge described in Sec.~\ref{sec:potential}.

\subsection{Functional Form}
\label{sec:phase_space_df}

A convenient functional form is a \textit{quasi-isothermal distribution function} (qDF), widely used for stellar disks \cite{Dehnen:1999ea}. We generalize it for FFPs by including an \textit{ejection kernel} $K(v)$ that captures the additional dispersion imparted at the time of ejection:

\begin{equation}
f(R,z;v_R,v_\phi,v_z)\propto
\nu(R,z)\,
\exp\!\left[
-\frac{v_R^2}{2\sigma_R^2(R)}
-\frac{\left(v_\phi-\overline{v_\phi}(R)\right)^2}{2\sigma_\phi^2(R)}
-\frac{v_z^2}{2\sigma_z^2(R)}
\right]
K(v).
\label{eq:kernelequation}
\end{equation}

Here:

\begin{itemize}
\item $\nu(R,z)$ is the FFP number density profile, assumed to trace (to leading order) the stellar disk density with scale length $R_d$ and scale height $z_d$.
\item $\sigma_i(R)$ are the velocity dispersions, declining with radius as in the stellar disk, e.g. $\sigma_i(R)\sim\sigma_{i,0}\exp[-(R-R_0)/R_\sigma]$.
\item $\overline{v_\phi}(R)$ encodes the mean rotation of the disk, corrected for asymmetric drift.
\item $K(v)$ is an additional factor capturing the contribution from ejection kicks.
\end{itemize}

We note that Eq.~\eqref{eq:kernelequation} models the ejection kernel as a multiplicative modification of the equilibrium disk DF rather than as a convolution of the parent (pre-kick) DF with the kick-velocity kernel; the latter would be the more rigorous treatment of how an impulsive kick reshapes a velocity distribution; we adopt the simpler multiplicative form here as a phenomenological placeholder, consistent with the "convenient functional form" framing in Sec.~\ref{sec:phase_space_df}, and flag the convolution treatment as a natural refinement for future work.

The ejection kernel $K(v)$ can be modeled as a multiplicative modification to the velocity distribution. Motivated by the low-velocity peak and high-velocity tail seen in $N$-body ejection-velocity distributions \citep{2025arXiv251204700A,ColemanDeRocco2025,2016MNRAS.461L.107M} (Sec.~\ref{sec:kicks}), a simple analytic form is:

\begin{equation}
K(v)\propto
\frac{v^2}{\left(v^2+v_0^2\right)^\alpha},
\label{eq:ejectionkernel}
\end{equation}

with $v=|\mathbf{v}-\mathbf{v}_\star|$ the velocity relative to the progenitor stellar motion, $v_0\sim1$--$3\ \mathrm{km\,s^{-1}}$ setting the scale, and $\alpha\gtrsim2$ controlling the high-velocity fall-off. Equation (\ref{eq:ejectionkernel}) ensures a low-velocity rise reminiscent of a Maxwell--Boltzmann distribution, with a suppressed power-law tail at large $v$.

\subsection{Discussion}

The distribution function of Eq.~(\ref{eq:kernelequation}) has several desirable properties:

\begin{enumerate}
\item It reduces to a standard anisotropic Gaussian ellipsoid in velocity space when $K(v)=1$, consistent with stellar disk kinematics.
\item It naturally incorporates an additional ejection-driven broadening of the velocity distribution through $K(v)$.
\item It is axisymmetric in position space, consistent with the smooth Galactic disk approximation, while still permitting vertical structure through $\nu(R,z)$.
\end{enumerate}

\subsection{Microlensing Event Rate Formalism}
\label{sec:MicroEventRateFormalism}

The microlensing event rate per source star depends on the phase-space distribution of lenses along the line of sight, together with the microlensing cross section for a given lens mass and geometry. For a lens with Einstein radius $R_E$ moving with transverse velocity $v_\perp$ relative to the source, the effective area swept per unit time is

\begin{equation}
\sigma_{\rm ml}=2R_Ev_\perp,
\label{eq:sigmaeq}
\end{equation}

for point-lens microlensing in the geometric limit, where
\begin{equation}
R_E = \sqrt{\frac{4GM}{c^2}\, \frac{D_L (D_S - D_L)}{D_S}}.
\end{equation}
Because $\sigma_{\rm ml}$ already carries the single power of $v_\perp$ associated with the sweep rate, the differential event rate for a population of FFPs with distribution function $f(\mathbf{x},\mathbf{v})$ is obtained by multiplying $\sigma_{\rm ml}$ directly by $f$ and integrating over the lens population, without an additional explicit factor of $v_\perp$:

\begin{equation}
\frac{d\Gamma}{dt_E}
=
\int_{0}^{D_S} dD_L
\int d^{3}v\;
f(\mathbf{x},\mathbf{v})\,
\sigma_{\rm ml}\,
\delta\!\left(t_E-\frac{R_E}{v_\perp}\right)
\label{eq:dGamma_dtE}
\end{equation}

where
\begin{itemize}
  \item $D_S$ is the distance to the source star (typically in the Galactic bulge),
  \item $D_L$ is the distance to the lens,
  \item $M$ is the lens mass,
  \item $v_\perp$ is the lens--source relative transverse velocity.
\end{itemize}

We stress this point because it is a common source of error: $\sigma_{\rm ml}=2R_Ev_\perp$ is not a true cross-sectional area to be multiplied by an independent flux factor of $v_\perp$, as in the textbook kinetic-theory rate $n\sigma v$ for a fixed-area target; it already \emph{is} the area-sweep rate (units of area per time), so including a further explicit $v_\perp$ alongside it would double count the lens's transverse motion.

\subsection{Reduction to the Transverse-Velocity Ellipsoid}
\label{sec:reduction}

Decompose the velocity into a line-of-sight component $v_\parallel$ and a two-dimensional transverse
velocity $(v_y,v_z)$, and write the transverse velocity in polar form,
$v_\perp=\sqrt{v_y^2+v_z^2}$, $\tan\varphi=v_z/v_y$, so that
\begin{equation}
d^3v = dv_\parallel\, v_\perp\, dv_\perp\, d\varphi,
\label{eq:polar_jacobian}
\end{equation}
the middle factor being the ordinary two-dimensional polar Jacobian.
Assuming the DF factorizes along the line of sight,
$f=\nu(\mathbf x)\,g_\parallel(v_\parallel)\,g_\perp(v_y,v_z)$, each
normalized to unity, the line-of-sight integral is trivial and drops
out -- consistent with the fact that only $\sigma_\phi(R)$ and
$\sigma_z(R)$, and not $\sigma_R(R)$, enter the transverse-velocity
ellipsoid used numerically in Sec.~\ref{sec:numerical_implementation}.
Substituting Eqs.~\eqref{eq:sigmaeq} and \eqref{eq:polar_jacobian} into
Eq.~\eqref{eq:dGamma_dtE}, the velocity integral carries two explicit
powers of $v_\perp$ -- one from $\sigma_{\rm ml}=2R_Ev_\perp$ and one
from the polar Jacobian -- giving
\begin{equation}
\frac{d\Gamma}{dt_E}
=
\int_0^{D_S} dD_L\;
2R_E\,\nu(\mathbf x)
\int d\varphi
\int dv_\perp\, v_\perp^2\, g_\perp(v_\perp,\varphi)\,
\delta\!\left(t_E-\frac{R_E}{v_\perp}\right).
\end{equation}
Using $\delta(t_E-R_E/v_\perp)=(R_E/t_E^2)\,\delta(v_\perp-R_E/t_E)$ and
evaluating $v_\perp^2\to(R_E/t_E)^2$ at the root gives an overall
prefactor $2R_E^4/t_E^4$ multiplying the bare angular integral
$\int d\varphi\,g_\perp(R_E/t_E,\varphi)$:
\begin{equation}
\frac{d\Gamma}{dt_E}
=
\int_0^{D_S} dD_L\;
\frac{2R_E^4}{t_E^4}\,
\nu(\mathbf x)
\int d\varphi\, g_\perp\!\left(\frac{R_E}{t_E},\varphi\right).
\label{eq:distributionmicro_angular}
\end{equation}
Numerically, we do not evaluate the bare angular integral
$\int d\varphi\,g_\perp$ directly; we instead evaluate the marginal
transverse-\emph{speed} density
\begin{equation}
Q(s) \equiv s\int_0^{2\pi} d\varphi\, g_\perp(s,\varphi),
\end{equation}
i.e.\ the same "extra factor of $s$ from polar coordinates" that turns
a two-dimensional Gaussian into a Rayleigh-type speed distribution.
Substituting
$\int d\varphi\,g_\perp(R_E/t_E,\varphi)=(t_E/R_E)\,Q(R_E/t_E)$
into Eq.~\eqref{eq:distributionmicro_angular} removes one
power of $R_E/t_E$ from the $2R_E^4/t_E^4$ prefactor, giving the final expression
\begin{equation}
\frac{d\Gamma}{dt_E}
=
\int_0^{D_S} dD_L\;
\frac{2R_E^3}{t_E^3}\,
\nu(\mathbf x)\,
Q\!\left(\frac{R_E}{t_E}\right),
\label{eq:distributionmicro}
\end{equation}
with $Q$ evaluated from the anisotropic Gaussian ellipsoid of Sec.~\ref{sec:symmetries}. This is the expression implemented in the numerical pipeline of Sec.~\ref{sec:numerical_implementation}.

For an axisymmetric DF of the form in Eq. (\ref{eq:kernelequation}), the velocity integral reduces to evaluating the probability of drawing a transverse velocity $v_\perp=R_E/t_E$ from the anisotropic Gaussian ellipsoid, modulated by the ejection kernel $K(v)$.

\subsection{Connection to Roman Predictions}
\label{sec:connectiontoRoman}

Roman's microlensing survey will monitor millions of bulge stars with sensitivity to short-timescale events. The relevant observable is the total event rate above a threshold magnification, given by

\begin{equation}
\Gamma=
\int
\mathrm{d}t_E\,
\frac{\mathrm{d}\Gamma}{\mathrm{d}t_E},
\label{eq:gammaeq}
\end{equation}

and the distribution $\mathrm{d}\Gamma/\mathrm{d}t_E$ itself, which determines the relative abundance of Earth-mass, Neptune-mass, and Jupiter-mass FFPs detectable by Roman. By inserting the DF from Eq. (\ref{eq:kernelequation}) into Eq. (\ref{eq:distributionmicro}), we obtain predictions for the expected number and timescale distribution of FFP microlensing events, which can be directly compared to Roman observations.

Thus the framework outlined here establishes a direct connection between ejection physics, Galactic phase-space evolution, and microlensing observables.

\section{Microlensing Implementation \& Results}
\label{sec:numerical_implementation}

\begin{figure} [H]
    \centering
    \includegraphics[width=1.0\linewidth]{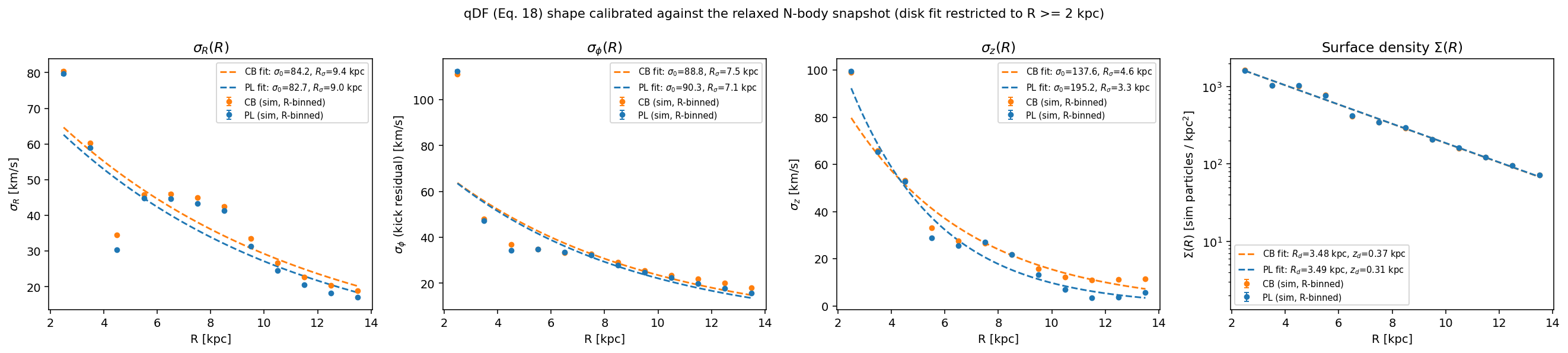}
    \caption{Gaussian velocity ansatz fit to the final-time N-body snapshot (still dynamically evolving; see Sec.~\ref{sec:nokick_control_results}), separately per channel, restricted to R>=2 kpc as described above.}
    \label{fig:qdf_calibration}
\end{figure}

We note that the theoretical ejection kernel $K(v)$ (Eq.~\ref{eq:ejectionkernel}) of Sec.~\ref{sec:phase_space_df} is a separate, illustrative analytic form and is not the mechanism by which mass-dependence enters the numerical pipeline of this section; here, the velocity ellipsoid's mass dependence is instead measured directly from the final-time N-body snapshot via the empirical scaling exponents $p_{\sigma_\phi}, p_{\sigma_z}$ (Sec.~\ref{sec:pipeline_output}), bypassing $K(v)$ entirely.

\subsection{Pipeline Output at the Adopted Geometry}
\label{sec:pipeline_output}

Applying the procedure of Sec.~\ref{sec:Microlensing}, we compute, for each channel and mass bin, a peak Einstein timescale, an unweighted $\Gamma$, and a mass-function-weighted $\Gamma$ (Table~\ref{tab:event_rate_summary}). 
 
\begin{table}[H]
\centering
\caption{ Earth-like, Neptune-like, and Jupiter-like correspond to the 1--3, 10--30, and 100--317~$M_\oplus$ simulation bins (Sec.~\ref{sec:mass_resolved_diagnostics}), respectively; the 3--10 and 30--100~$M_\oplus$ bins are not separately propagated through the microlensing pipeline.}
\label{tab:event_rate_summary}
\begin{tabular}{llccc}
\hline
Channel & Mass bin & $t_{E,\rm peak}$ [days] & $\Gamma_{\rm unweighted}$ & $\Gamma_{\rm weighted}$ \\
\hline
CB & Earth-like    & 0.0390 & 1.73 & 0.767 \\
CB & Neptune-like  & 0.0739 & 5.48 & 0.212 \\
CB & Jupiter-like  & 0.2261 & 56.30 & 0.021 \\
PL & Earth-like    & 0.0390 & 1.73 & 0.342 \\
PL & Neptune-like  & 0.0739 & 5.48 & 0.265 \\
PL & Jupiter-like  & 0.2261 & 56.30 & 0.393 \\
\hline
\end{tabular}
\end{table}

\begin{figure}[H]
    \centering
    \includegraphics[width=1.0\linewidth]{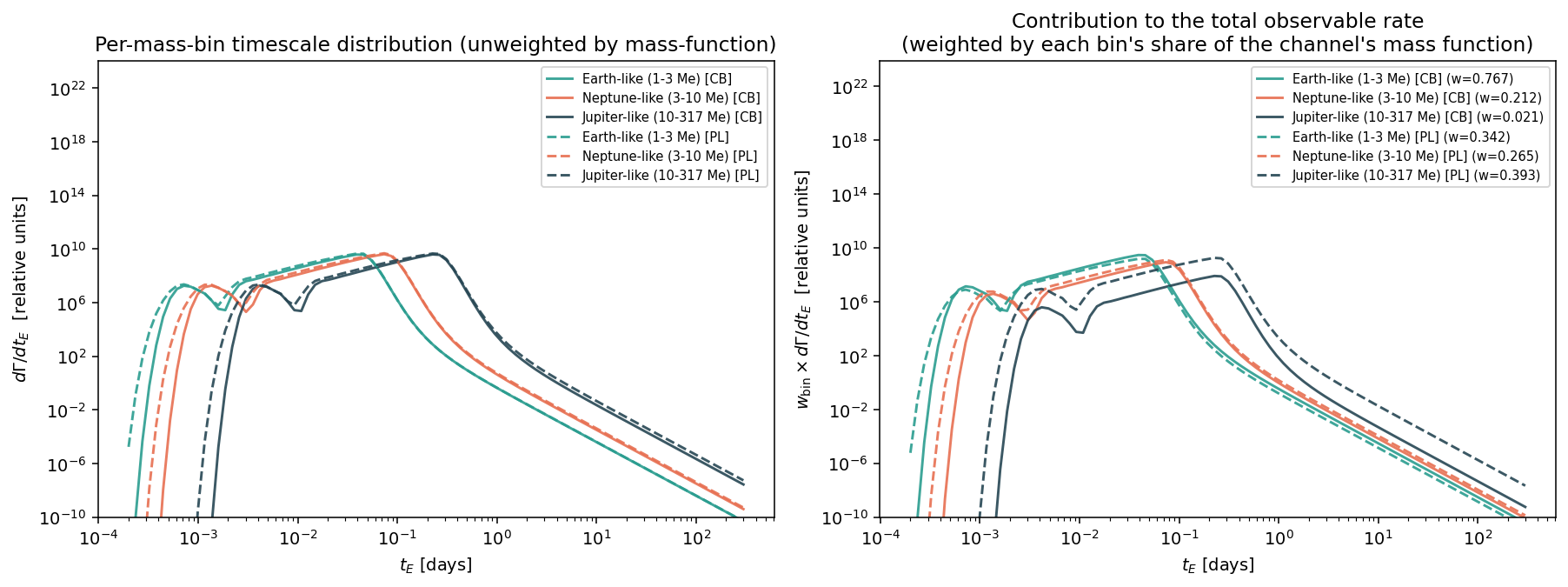}
    \caption{Left: ${\rm d}\Gamma/{\rm d}t_E$ per channel and mass bin,
unweighted. Right: same, weighted by each bin's share of its channel's
sampled mass function. Illustrative only; see scope note, Sec.~\ref{sec:Microlensing}.}
    \label{fig:microlensing_timescales}
\end{figure}

\begin{figure}[H]
    \centering
    \includegraphics[width=0.5\linewidth]{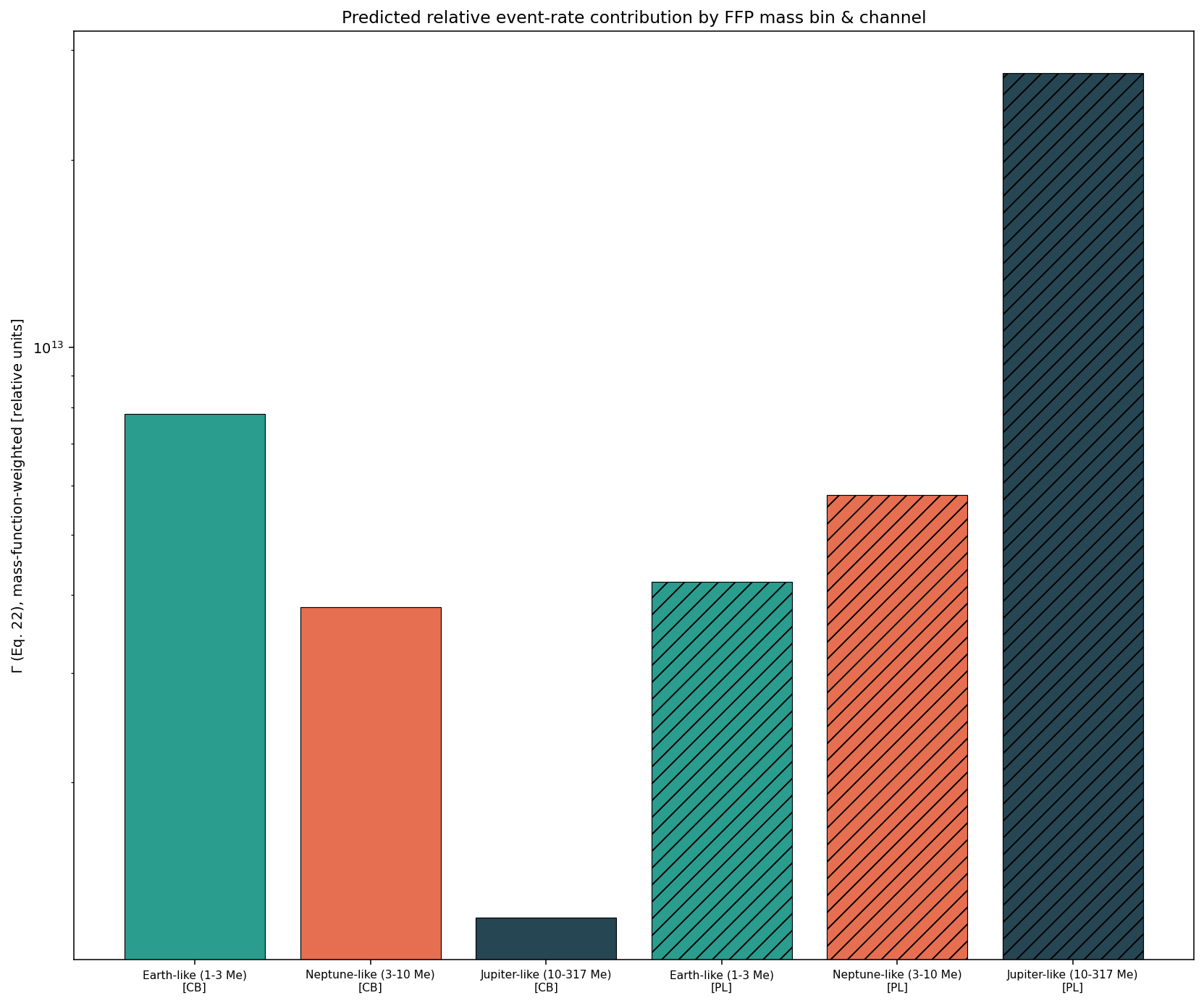}
    \caption{Mass-function-weighted $\Gamma$ by
channel and mass bin. Illustrative only; see scope note, Sec.~\ref{sec:Microlensing}.}
    \label{fig:microlensing_relative_abundance}
\end{figure}

The pipeline output summarized in Table \ref{tab:event_rate_summary} and Figures \ref{fig:microlensing_timescales}--\ref{fig:microlensing_relative_abundance} provides a schematic, shape-level connection between the ejection and Galactic-dynamics results of Sections \ref{sec:kicks}--\ref{sec:results} and the microlensing observable of Eq.~(\ref{eq:dGamma_dtE})--(\ref{eq:gammaeq}). We emphasize at the outset, as already flagged in the table and figure captions, that these results are illustrative: the quantities plotted are built from Eq.~(\ref{eq:distributionmicro}) using a number density $\nu_0(R)$ fit directly to the simulated particle counts, with no external normalization to a physical FFP number density, source-star count, survey exposure time, or detection efficiency. Table \ref{tab:event_rate_summary} and Figures \ref{fig:microlensing_timescales}--\ref{fig:microlensing_relative_abundance} should therefore be read as comparisons of \emph{relative shape} across mass bins and channels, not as absolute predicted Roman event rates.

As expected from $R_E\propto\sqrt{M}$ (Sec.~\ref{sec:MicroEventRateFormalism}), the peak Einstein timescale increases monotonically with lens mass in both channels: $t_{E,\mathrm{peak}}=0.0390$, $0.0739$, and $0.226$ days for the Earth-like, Neptune-like, and Jupiter-like bins, respectively (Table \ref{tab:event_rate_summary}). All three values remain squarely in the sub-day regime that motivates Roman's high-cadence bulge monitoring strategy \citep{Penny2019,Johnson2020}. We note that the Earth-to-Neptune step in $t_{E,\mathrm{peak}}$ ($1.9\times$) falls short of the $\sim\sqrt{10}\approx3.2\times$ expected from $R_E\propto\sqrt{M}$ over the corresponding $\sim10\times$ mass step, while the Neptune-to-Jupiter step ($3.06\times$) matches this scaling well; given the identical, grid-resolution origin of the CB/PL degeneracy discussed next, we attribute this shortfall to the same fixed 90-point logarithmic $t_E$ grid rather than to a genuine departure from $R_E\propto\sqrt{M}$, but we have not verified this quantitatively and flag it as a check worth performing with a finer grid. As before, $t_{E,\mathrm{peak}}$ is identical between the CB and PL channels within each mass bin to the precision quoted; because $t_{E,\mathrm{peak}}$ is read off the same fixed 90-point logarithmically spaced grid described above, this remains at least partly a grid-resolution artifact rather than evidence that the two channels' transverse-velocity ellipsoids peak at literally the same timescale.

\textbf{Channel-independence of $\Gamma_{\rm unweighted}$.} The unweighted event rate $\Gamma_{\rm unweighted}$ in Table \ref{tab:event_rate_summary} is likewise identical between the CB and PL channels within each mass bin. Unlike $t_{E,\mathrm{peak}}$, this is not a grid artifact: $\Gamma_{\rm unweighted}$ depends on the Einstein radius $R_E\propto\sqrt{M}$, which is set only by the (shared) physical mass bin, and on the final-time transverse-velocity ellipsoid, whose CB and PL parameters differ by only $\sim1\%$ at the final snapshot (Table \ref{tab:final_velocity_comparison}) -- below the precision quoted in Table \ref{tab:event_rate_summary}. The two channels are therefore not literally forced to agree in $\Gamma_{\rm unweighted}$; they agree numerically because their underlying velocity ellipsoids are already nearly identical by this point in the integration (Sec.~\ref{sec:results_velocity}). Only $\Gamma_{\rm weighted}$, which folds in each channel's distinct sampled mass function, differs between CB and PL.

\section{Conclusion}
\label{sec:conclusion}

We have carried out a controlled numerical experiment testing whether mass-dependent ejection velocities leave a mass-dependent signature in the Galactic phase-space distribution of free-floating planets. We evolved $10^6$ collisionless test particles for $10^8$~yr in a static, phenomenological Milky Way potential, comparing mass-independent and explicitly mass-coupled kick prescriptions for circumbinary (CB) and planet--planet scattering (PL) channels. A matched no-kick control population was used to separate the incremental effect of the ejection kick from the subsequent dynamical evolution of the initially warm population.

The coupled model produces the intended qualitative mass dependence at injection: the sampled median kick speed decreases with increasing mass in both ejection channels. For the combined population, the median kick decreases from $10.716$ to $0.708~{\rm km\,s^{-1}}$ across the five mass bins, corresponding to a factor of $\sim15.1$ range (Table~\ref{tab:mass_binned_kicks}). The sampled values follow the expected declining trend, although the highest-mass PL bins show some flattening relative to the idealized $M^{-1/2}$ scaling. This confirms that the intended mass--velocity correlation is present in the injected population, while also illustrating the importance of checking the realized sampling distribution against the analytic prescription.

The injected hierarchy is reflected in the matched kicked-versus-control displacement after $100$~Myr. In the coupled model, the median displacement decreases from $0.7001$ to $0.0539$~kpc across the five mass bins, whereas the corresponding null-model values decrease from $0.7687$ to $0.3329$~kpc (Table~\ref{tab:displacement_mass_resolved}). The coupled model therefore exhibits a substantially stronger mass-dependent displacement trend. However, the null model also shows mass-dependent displacement despite its approximately mass-independent kick prescription. The spatial diagnostic should consequently be interpreted as a combined result of the injected kick distribution, the sampled initial conditions, and subsequent orbital evolution in the Galactic potential, with the stronger coupled-model trend being consistent with the imposed mass--kick correlation.

The contrast is weaker in the bulk kinematic moments. The final velocity dispersions of the null and coupled models are nearly indistinguishable, with combined values of $\sigma_R\simeq87.7$ and $87.6~{\rm km\,s^{-1}}$ and $\sigma_z\simeq82.1$ and $82.0~{\rm km\,s^{-1}}$, respectively (Table~\ref{tab:final_velocity_comparison}). The matched-control comparison gives kick-attributable quadrature excesses of only $\sim9~{\rm km\,s^{-1}}$ in $\sigma_R$ and $\sim5~{\rm km\,s^{-1}}$ in $\sigma_z$ for the combined population (Table~\ref{tab:control_comparison}). These results indicate that the global velocity statistics are dominated by the evolution of the initially warm population in the Galactic potential, while the incremental effect of the kick is comparatively modest. The mass-dependent ejection prescription is therefore more clearly reflected in the resolved matched spatial displacement than in the final global velocity moments.

As a supplementary settling diagnostic, we also evolved a reduced $N=10^4$ no-kick population for $10^9$~yr using a coarser timestep of $\Delta t=10^5$~yr. Its vertical velocity dispersion increased from $\sim32~{\rm km\,s^{-1}}$ initially to $\sim76~{\rm km\,s^{-1}}$ at $500$~Myr and $\sim79~{\rm km\,s^{-1}}$ at $1$~Gyr, suggesting that the rapid evolution of the initial non-equilibrium state becomes less pronounced at late times. Because of the reduced particle number and coarser timestep, this run is not used to infer a reliable long-term kick signature, nor does it establish complete dynamical equilibration. A robust assessment of the survival of mass-dependent kick signatures at Gyr timescales will require higher-resolution integrations, convergence tests, and a self-consistent equilibrium initial distribution function.

Finally, the microlensing calculation provides only a schematic, shape-level illustration of how the simulated phase-space distribution enters an observable. Because it is based on the same non-equilibrium final snapshot and lacks an absolute FFP-density normalization, source-star population, survey exposure, and detection efficiency, it should not be interpreted as an absolute prediction for the Roman event yield. The present calculation therefore establishes the relative contributions of injected kick effects and background dynamical evolution within the adopted numerical setup. A physically general prediction for the present-day Galactic FFP population will require future simulations with a dynamically self-consistent initial distribution, longer propagation times, improved numerical validation, and a fully normalized microlensing survey model.

\appendix
\section{Summary of Adopted Parameter Values}
\label{app:parameters}
 
This appendix collects the numerical parameter values used in the Galactic FFP simulations and the associated microlensing calculation. Unless otherwise stated, the fiducial values listed below correspond to the simulations used to generate the results presented in the main text. Quantities described as convergence tests are numerical diagnostics performed separately from the fiducial production run.
 
\subsection{Kick Velocity Distribution: Circumbinary (CB) Channel}
\label{app:kick_cb}
 
\begin{table}[H]
\centering
\caption{Adopted parameters for the circumbinary (CB) ejection channel.}
\label{tab:appendix_cb}
\begin{tabular}{lcl}
\toprule
\textbf{Parameter} & \textbf{Symbol} & \textbf{Value} \\
\midrule
Null-model kick dispersion & $\sigma_{\rm CB}$ & $10~{\rm km\,s^{-1}}$ \\
Mass-coupling exponent & & $-0.15$ \\
Mass range & $M$ & $1$--$317\,M_\oplus$ \\
\bottomrule
\end{tabular}
\end{table}
 
In the null model, the three Cartesian components of the CB kick are drawn independently from Gaussian distributions with standard deviation $\sigma_{\rm CB}=10~{\rm km\,s^{-1}}$. In the coupled model, the dispersion is modified according to
\begin{equation}
\sigma_{\rm CB,eff}(M)
=
10
\left(\frac{M}{M_\oplus}\right)^{-0.15}
{\rm km\,s^{-1}}.
\end{equation}
The exponent is an adopted modeling prescription rather than a universal theoretical scaling.
 
\subsection{Kick Velocity Distribution: Planet--Planet Scattering (PL) Channel}
\label{app:kick_pl}
 
\begin{table}[H]
\centering
\caption{Adopted parameters for the planet--planet scattering (PL) ejection channel.}
\label{tab:appendix_pl}
\begin{tabular}{lcl}
\toprule
\textbf{Parameter} & \textbf{Symbol} & \textbf{Value} \\
\midrule
Null-model kick-speed scale & $v_k$ & $4~{\rm km\,s^{-1}}$ \\
Null-model Gaussian width & & $1~{\rm km\,s^{-1}}$ \\
Mass-coupling exponent & & $-0.5$ \\
Mass range & $M$ & $1$--$317\,M_\oplus$ \\
\bottomrule
\end{tabular}
\end{table}
 
For the null model, the PL kick speed is drawn from a narrow Gaussian distribution centered on $4~{\rm km\,s^{-1}}$ with a width of $1~{\rm km\,s^{-1}}$. In the coupled model, the characteristic PL kick speed follows
\begin{equation}
v_{\rm PL,base}(M)
=
4
\left(\frac{M}{M_\oplus}\right)^{-1/2}
{\rm km\,s^{-1}}.
\end{equation}
As for the CB prescription, the exponent is an adopted model choice used to test whether a mass-dependent ejection velocity can survive subsequent Galactic propagation.
 
\subsection{Shared Mass Sampling and Channel Mixture}
\label{app:kick_shared}
 
\begin{table}[H]
\centering
\caption{Mass-sampling and channel-mixture parameters used in the simulations.}
\label{tab:appendix_shared}
\begin{tabular}{lc}
\toprule
\textbf{Parameter} & \textbf{Value} \\
\midrule
Sampled mass range & $1$--$317\,M_\oplus$ \\
Underlying mass function & $dN/dM\propto M^{-1.3}$ \\
CB mass-function turnover & $[1+(M/3M_\oplus)^2]^{-1}$ \\
CB fraction & $F_{\rm CB}=0.5$ \\
PL fraction & $1-F_{\rm CB}=0.5$ \\
\bottomrule
\end{tabular}
\end{table}
 
The CB and PL populations are sampled independently according to their respective mass distributions and then combined with a fixed channel fraction of $F_{\rm CB}=0.5$. The mass distributions therefore determine the number of objects in each mass bin, while the kick prescription determines the mass dependence of the ejection velocity.
 
We emphasize that $F_{\rm CB}=0.5$ is an equal-weight choice adopted for this controlled dynamical experiment, and is independent of the physically motivated abundance normalizations $\eta_{\rm CB}\sim0.1$--$0.5$ planets per binary and $\eta_{\rm ej}\sim2$--$8$ planets per star introduced in Sec.~\ref{sec:Source_term}. Those normalizations set the physical number density scale of each channel in a realistic Galactic FFP population, whereas $F_{\rm CB}$ only sets the relative sample size of the two channels within our simulation, chosen so that both are populated with comparable statistical precision for the kicked-vs-control comparison. A population-synthesis calculation aiming to predict the realistic relative abundance of CB- and PL-produced FFPs in the Galaxy -- rather than to test, in a controlled way, whether a mass-dependent kick survives Galactic propagation -- would instead need to weight the two channels by $\eta_{\rm CB}$ and $\eta_{\rm ej}$, which could favor the PL channel by a large factor depending on where in each adopted range the true values lie; we leave this reweighting to future work.
 
\subsection{Initial-Condition Sampling}
\label{app:initial}
 
\begin{table}[H]
\centering
\caption{Initial spatial and kinematic distributions used for the warm-disk FFP population.}
\label{tab:appendix_initial}
\begin{tabularx}{\textwidth}{lclX}
\toprule
\textbf{Quantity} & \textbf{Distribution} & \textbf{Parameter} & \textbf{Description} \\
\midrule
Cylindrical radius $R$
& Gamma
& $k=2,\ \theta=R_d$
& $R_d=3.5$~kpc \\
 
Azimuth $\phi$
& Uniform
& $(0,2\pi)$
& Isotropic in azimuth \\
 
Vertical height $z$
& Laplace
& $z_d=0.3$~kpc
& Symmetric about the Galactic plane \\
 
Radial velocity $v_R$
& Gaussian
& $\sigma_R(R)$
& Radius-dependent warm-disk dispersion \\
 
Vertical velocity $v_z$
& Gaussian
& $\sigma_z(R)$
& Radius-dependent warm-disk dispersion \\
 
Azimuthal velocity $v_\phi$
& Gaussian
& $0.7\sigma_R$
& Centered on the asymmetric-drift-corrected mean rotation \\
 
\bottomrule
\end{tabularx}
\end{table}
 
The radial and vertical velocity components are drawn as
\begin{equation}
v_R\sim\mathcal{N}(0,\sigma_R),
\qquad
v_z\sim\mathcal{N}(0,\sigma_z),
\end{equation}
while the azimuthal component is drawn about the asymmetric-drift-corrected mean rotation velocity. The resulting population is therefore a warm rotating disk rather than a cold population initialized exactly on circular orbits.
 
For every kicked particle, a matched no-kick control twin is initialized with the same birth position, mass, and pre-kick Galactic velocity. The only difference between the kicked particle and its control twin is the addition of the ejection kick. This construction allows the subsequent evolution of the initial warm population to be separated from the incremental effect of the kick.
 
\subsection{Galactic Potential}
\label{app:potential}
 
The Galactic potential is the sum of disk, triaxial bulge, and NFW-like halo components,
\begin{equation}
\Phi_{\rm tot}
=
\Phi_{\rm disk}
+
\Phi_{\rm bulge}
+
\Phi_{\rm halo}.
\end{equation}
 
The adopted potential parameters are given in Tables~\ref{tab:disk_parameters}, \ref{tab:bulge_parameters}, and \ref{tab:halo_parameters} in the main text. The halo regularization parameter $R_{0,\rm halo}$ is used only as a numerical guard near $r=0$ and is distinct from the finite-difference force-evaluation scale.
 
\subsection{Numerical Integration and Force Evaluation}
\label{app:integration}
 
\begin{table}[H]
\centering
\caption{Numerical parameters used for the fiducial Galactic integrations.}
\label{tab:appendix_integration}
\begin{tabularx}{\textwidth}{Xc}
\toprule
\textbf{Parameter} & \textbf{Fiducial value} \\
\midrule
Total propagation time $T$
& $10^{8}$~yr \\
 
Timestep $\Delta t$
& $10^{4}$~yr \\
 
Number of integration steps
& $10{,}000$ \\
 
Particles per production run
& $1{,}000{,}000$ \\
 
CB particles
& $500{,}000$ \\
 
PL particles
& $500{,}000$ \\
 
CB fraction $F_{\rm CB}$
& $0.5$ \\
 
Finite-difference step $h$
& $10^{-2}$~kpc \\
 
Radial safeguard $R_{\rm eps}$
& $10^{-3}$~kpc \\
 
Integrator
& Kick--Drift--Kick leapfrog \\
 
Potential
& Static disk + triaxial bulge + NFW-like halo \\
 
Stored snapshots
& $t=0$, $5\times10^{7}$~yr, $10^{8}$~yr \\
 
\bottomrule
\end{tabularx}
\end{table}
 
The acceleration is calculated using a centered finite-difference approximation to the gradient of the total potential,
\begin{equation}
a_x
\simeq
-\frac{\Phi(x+h,y,z)-\Phi(x-h,y,z)}{2h},
\end{equation}
with analogous expressions for $a_y$ and $a_z$. The fiducial calculation uses $h=0.01$~kpc.
 
 
A force-resolution convergence test is \emph{designed} as follows: 
\begin{equation}
h =
0.005,\ 0.01,\ 0.02,\ 0.04~{\rm kpc},
\end{equation}
using the same initial particle realization for each value of $h$ so that differences between runs arise from the numerical force evaluation rather than Monte Carlo sampling, and compare the final-time radial and vertical velocity dispersions, $\sigma_R$ and $\sigma_z$, and the disk-departure fraction $f_{\rm disk,dep}$, via the maximum fractional variation
\begin{equation}
\delta_{\rm max}
=
\max_h
\left[
\frac{|X(h)-X(h_{\rm fid})|}
{|X(h_{\rm fid})|}
\right],
\end{equation}
where $X$ denotes each of $\sigma_R$, $\sigma_z$, and $f_{\rm disk,dep}$ and $h_{\rm fid}=0.01$~kpc. We report this methodology in full so that the test is straightforward for us, or an independent party, to carry out and check against the fiducial results; until it is run, the numerical convergence of the fiducial $h=0.01$~kpc choice should be regarded as plausible (Sec.~\ref{sec:acceleration}) but not verified, and Table~\ref{tab:appendix_integration} lists $h$ among the fiducial parameters without an accompanying $\delta_{\rm max}$ value.

The leapfrog integrator is applied in kick--drift--kick form in the fixed Galactic potential. No intrinsic distribution of FFP ejection times is sampled: the ejection kick is applied at the initial time, $t=0$, and the subsequent integration represents post-ejection Galactic propagation.
 
\subsection{Matched No-Kick Control: Definition}
\label{app:control}
 
This subsection defines the construction of the control population; the resulting kicked-vs-control comparison is presented in Sec.~\ref{sec:nokick_control_results}. The matched control population is constructed from the same initial conditions as the kicked population. For each particle, the control retains the identical birth position, mass, and pre-kick velocity but receives no ejection velocity. Both the kicked and control populations are then propagated through the same static Galactic potential using the same integration scheme.
 
Kick-attributable velocity-dispersion excesses are quantified through the quadrature difference
\begin{equation}
\sigma_{\rm excess}
=
\sqrt{
\max
\left(
0,
\sigma_{\rm kick}^2-\sigma_{\rm control}^2
\right)
}.
\label{eq:sigma_excess}
\end{equation}
 
This quantity is used to distinguish velocity dispersion generated by the initial warm-disk population and its subsequent Galactic evolution from the additional contribution associated with the ejection kick.
 
\subsection{Escape and Disk-Departure Diagnostics}
\label{app:diagnostics}
 
\begin{table}[H]
\centering
\caption{Thresholds used for the Galactic escape and disk-departure diagnostics.}
\label{tab:appendix_diagnostics}
\begin{tabular}{lc}
\toprule
\textbf{Quantity} & \textbf{Value} \\
\midrule
Disk-departure height $Z_{\rm escape}$ & $2$~kpc \\
Galactic unbound condition & $E_i>0$ \\
\bottomrule
\end{tabular}
\end{table}
 
The instantaneous specific orbital energy is
\begin{equation}
E_i
=
\frac{1}{2}v_i^2+\Phi_{\rm tot}(\mathbf{x}_i).
\end{equation}
Particles with $E_i>0$ are classified as gravitationally unbound from the adopted Galactic potential.
 
Separately, the disk-departure fraction is defined as
\begin{equation}
f_{\rm disk,dep}(M,t)
=
P\left(|z|>2~{\rm kpc}\mid M,t\right).
\end{equation}
This threshold measures departure from the adopted disk region and is not equivalent to Galactic escape.
 
\subsection{Phase-Space KDE Settings}
\label{app:kde}
 
\begin{table}[H]
\centering
\caption{Kernel-density estimation settings used for the phase-space visualizations.}
\label{tab:appendix_kde}
\begin{tabular}{lc}
\toprule
\textbf{Parameter} & \textbf{Value} \\
\midrule
Grid resolution & $120\times120$ \\
Maximum particles subsampled per channel & $50{,}000$ \\
\bottomrule
\end{tabular}
\end{table}
 
The KDE calculations are used for visualization of the simulated phase-space distributions. They do not modify the underlying particle trajectories or enter the calculation of the reported global velocity-dispersion statistics.
 
\subsection{Propagation Diagnostics}
\label{app:propagation}
 
The propagation analysis evaluates the net displacement of each FFP relative to its birth position,
\begin{equation}
\Delta r_i(t)
=
\left|
\mathbf r_i(t)-\mathbf r_i(0)
\right|
\end{equation}
or, explicitly,
\begin{equation}
\Delta r_i(t)
=
\sqrt{
[x_i(t)-x_i(0)]^2+
[y_i(t)-y_i(0)]^2+
[z_i(t)-z_i(0)]^2
}.
\end{equation}
 
The median and upper-percentile displacement statistics are evaluated in the five mass bins used throughout the analysis. These quantities represent net displacement from the birth location, not the total path length traveled along the Galactic orbit.
 
Propagation snapshots are evaluated at
\begin{equation}
t=1,\ 10,\ 30,\ 100~{\rm Myr},
\end{equation}
where these times denote post-ejection propagation times rather than an assumed distribution of ejection times or planetary-system lifetimes.
 
\subsection{Mass Bins}
\label{app:massbins}
 
\begin{table}[H]
\centering
\caption{Mass bins used for all mass-resolved diagnostics.}
\begin{tabular}{cc}
\toprule
\textbf{Bin} & \textbf{Mass range} \\
\midrule
1 & $[1,3)\,M_\oplus$ \\
2 & $[3,10)\,M_\oplus$ \\
3 & $[10,30)\,M_\oplus$ \\
4 & $[30,100)\,M_\oplus$ \\
5 & $[100,317]\,M_\oplus$ \\
\bottomrule
\end{tabular}
\label{tab:appendix_massbins}
\end{table}
 
\subsection{Microlensing Pipeline Geometry and Grids}
\label{app:microlensing}
 
\begin{table}[H]
\centering
\caption{Fixed geometry and grid settings for the illustrative microlensing calculation. The observer radius $R_0$ and source distance $D_S$ are set to the same value because the source is placed at the Galactic Center itself along the adopted line of sight, rather than being drawn from a realistic bulge source-distance distribution; see the scope note below.}
\label{tab:appendix_microlensing}
\begin{tabular}{lc}
\toprule
\textbf{Parameter} & \textbf{Value} \\
\midrule
Observer Galactocentric radius $R_0$ & $8.2$~kpc \\
Source distance $D_S$ & $8.2$~kpc \\
Line of sight & Fixed toward the Galactic Center \\
$t_E$ grid & 90-point log-spaced, $2\times10^{-4}$--$300$~days \\
\bottomrule
\end{tabular}
\end{table}
 
The microlensing calculation is applied to the simulated FFP phase-space distribution as an illustrative mapping between the dynamical population and event-timescale distributions. Because it uses a fixed sightline and a single fixed source distance, and does not impose an absolute FFP density, source-star population, or survey selection function, the resulting quantities should be interpreted as relative-shape diagnostics rather than absolute survey yield predictions.
 
\subsection{Summary of Numerical Scope}
\label{app:scope}
 
The fiducial simulations therefore consist of $10^6$ FFP test particles propagated for $10^8$~yr with a timestep of $10^4$~yr in a static Galactic potential. Each FFP receives a single ejection kick at $t=0$, with either mass-independent or mass-dependent kick parameters. The matched no-kick population is evolved from the same pre-kick initial conditions.
 
The principal controlled comparison is consequently
\begin{equation}
M_{\rm FFP}
\longrightarrow
v_{\rm kick}(M)
\longrightarrow
\Delta r(M,t)
\longrightarrow
P(R,z\mid M,t),
\end{equation}
with the no-kick population providing the baseline against which the incremental dynamical effect of the kick is evaluated.
 
The simulation does not explicitly model the lifetime of the planetary system before ejection, an intrinsic distribution of ejection times, cluster dissolution, or a time-dependent Galactic potential. The results should therefore be interpreted as a controlled post-ejection propagation experiment under the stated initial-condition and potential assumptions.
 
\subsection{Supplementary $N=10^4$, 1 Gyr Integration}
\label{app:supplementary_1gyr}
 
Table~\ref{tab:appendix_1gyr} reports the full combined-channel (ALL) diagnostics for the supplementary integration referenced in Sec.~\ref{sec:conclusion}: $N=10^4$ particles (5000 CB, 5000 PL), $T=10^9$~yr, $\Delta t=10^5$~yr ($10^4$ steps), with snapshots at $t=0$, $5\times10^8$~yr (midway), and $10^9$~yr (final). The no-kick control is identical for the null and coupled runs, since it receives no kick.
 
\begin{table}[H]
\centering
\caption{Combined-channel (ALL) diagnostics for the supplementary $N=10^4$, $T=10^9$~yr integration. $\sigma_R$ and $\sigma_z$ are in km\,s$^{-1}$; $f_{\rm disk,dep}$ and $f_{\rm gal,esc}$ are fractions. $\Delta\sigma_R$, $\Delta\sigma_z$ are the kicked-vs-control quadrature excesses at the final snapshot (Eq.~\ref{eq:sigma_excess}); given the reduced particle count these are shown for completeness only and are \emph{not} used as evidence of a kick-induced signal (see Sec.~\ref{sec:conclusion}).}
\label{tab:appendix_1gyr}
\begin{tabular}{lccc}
\toprule
\textbf{Population} & \textbf{Initial ($t=0$)} & \textbf{Midway (500 Myr)} & \textbf{Final (1 Gyr)} \\
\midrule
\multicolumn{4}{l}{\textit{No-kick control (shared by null and coupled runs)}} \\
$\sigma_R$            & 53.70 & 86.52 & 85.96 \\
$\sigma_z$            & 31.81 & 76.45 & 78.74 \\
$f_{\rm disk,dep}$    & 0.001 & 0.034 & 0.044 \\
$f_{\rm gal,esc}$     & 0.0000 & 0.0006 & 0.0009 \\
\midrule
\multicolumn{4}{l}{\textit{Kicked, null model}} \\
$\sigma_R$            & 54.22 & 86.92 & 87.12 \\
$\sigma_z$            & 32.73 & 77.50 & 80.84 \\
$f_{\rm disk,dep}$    & 0.001 & 0.040 & 0.050 \\
$f_{\rm gal,esc}$     & 0.0000 & 0.0001 & 0.0008 \\
$\Delta\sigma_R$ (final) & \multicolumn{3}{c}{14.1} \\
$\Delta\sigma_z$ (final) & \multicolumn{3}{c}{18.3} \\
\midrule
\multicolumn{4}{l}{\textit{Kicked, coupled model}} \\
$\sigma_R$            & 54.11 & 87.76 & 87.75 \\
$\sigma_z$            & 32.54 & 78.03 & 79.56 \\
$f_{\rm disk,dep}$    & 0.001 & 0.040 & 0.050 \\
$f_{\rm gal,esc}$     & 0.0000 & 0.0008 & 0.0017 \\
$\Delta\sigma_R$ (final) & \multicolumn{3}{c}{17.6} \\
$\Delta\sigma_z$ (final) & \multicolumn{3}{c}{11.4} \\
\bottomrule
\end{tabular}
\end{table}
 
Two features of this run limit how much can be read into it beyond the control's own dispersion growth (Sec.~\ref{sec:conclusion}). First, at $N=10^4$ split evenly across two channels and five mass bins, the highest-mass CB bin ($100$--$317\,M_\oplus$) contains a single particle (compared to $N=56$ in the fiducial $10^6$-particle run), so no mass-resolved kick-speed or disk-departure statistics are reported for this run. Second, the final-time kicked-versus-control excesses shown above ($\Delta\sigma_R\sim14$--$18$, $\Delta\sigma_z\sim11$--$18~{\rm km\,s^{-1}}$) are comparable in magnitude to, and in one channel-resolved case (Table~\ref{tab:control_comparison}) larger than, the corresponding excesses from the fiducial $10^6$-particle, $10^8$~yr run; given the two-orders-of-magnitude smaller sample and the correspondingly larger standard error on each $\sigma$ ($\sim\sigma/\sqrt{2(N-1)}$), this is consistent with sampling noise rather than a genuine long-timescale kick signal, and we do not interpret it further. The disk-departure and Galactic-escape fractions remain at the few-percent level and sub-percent level, respectively, consistent with the fiducial run.

\bibliographystyle{JHEP}
\bibliography{bib}

@article{BhaskarPerets2025,
  author       = {Bhaskar, Hareesh Gautham and Perets, Hagai B.},
  title        = {Properties of Free-floating Planets Ejected through Planet-Planet Scattering},
  journal      = {The Astrophysical Journal},
  volume       = {991},
  number       = {2},
  pages        = {132},
  year         = {2025},
  month        = oct,
  eprint       = {2501.13166},
  archivePrefix= {arXiv},
  primaryClass = {astro-ph.EP},
  doi          = {10.3847/1538-4357/adf4e2}
}

@article{ZhaiLee2025,
  author       = {Zhai, Ruocheng and Lee, Man Hoi and Gan, Tianjun and Mao, Shude},
  title        = {Dynamical Instability of Multi-planet Systems and Free-floating Planets},
  journal      = {The Astrophysical Journal},
  volume       = {990},
  number       = {1},
  pages        = {74},
  year         = {2025},
  month        = aug,
  eprint       = {2507.21216},
  archivePrefix= {arXiv},
  doi          = {10.3847/1538-4357/adf557}
}

@article{VerasRaymond2012,
  author       = {Veras, Dimitri and Raymond, Sean N.},
  title        = {Planet–Planet Scattering Alone Cannot Explain the Free-Floating Planet Population},
  journal      = {MNRAS Letters},
  year         = {2012},
  volume       = {421},
  number       = {1},
  pages        = {L117--L121},
  eprint       = {1201.2175},
  archivePrefix= {arXiv},
  doi          = {10.1111/j.1745-3933.2012.01218.x}
}

@article{YuLai2024,
  author       = {Yu, Fangyuan and Lai, Dong},
  title        = {Free-floating Planets, Survivor Planets, Captured Planets, and Binary Planets from Stellar Flybys},
  journal      = {The Astrophysical Journal},
  volume       = {970},
  number       = {1},
  pages        = {97},
  year         = {2024},
  month        = jul,
  eprint       = {2403.07224},
  archivePrefix= {arXiv},
  doi          = {10.3847/1538-4357/ad4f81}
}

@article{Pfalzner2021,
  author       = {Pfalzner, S. and others},
  title        = {Significant Interstellar Object Production by Close Stellar Flybys},
  journal      = {Astronomy \& Astrophysics},
  year         = {2021},
  volume       = {651},
  pages        = {A38},
  doi          = {10.1051/0004-6361/202140587}
}

@article{Sumi2011,
  author       = {Sumi, Takahiro and Kamiya, Kazumi and Bennett, David P. and others},
  title        = {Unbound or distant planetary mass population detected by gravitational microlensing},
  journal      = {Nature},
  volume       = {473},
  pages        = {349--352},
  year         = {2011},
  doi          = {10.1038/nature10092}
}

@article{Mróz2017,
  author       = {Mróz, Przemek and Udalski, Andrzej and Skowron, Jan and others},
  title        = {No large population of unbound or wide-orbit Jupiter-mass planets},
  journal      = {Nature},
  volume       = {548},
  pages        = {183--186},
  year         = {2017},
  doi          = {10.1038/nature23276}
}

@article{Mróz2018,
  author       = {Mróz, Przemek and others},
  title        = {A Neptune-mass Free-floating Planet Candidate Discovered by Microlensing Surveys},
  journal      = {AJ},
  volume       = {155},
  pages        = {121},
  year         = {2018},
  doi          = {10.3847/1538-3881/aaaae9}
}

@article{Mróz2019,
  author       = {Mróz, Przemek and others},
  title        = {A terrestrial-mass rogue planet candidate detected in the shortest-timescale microlensing event},
  journal      = {ApJL},
  volume       = {903},
  pages        = {L11},
  year         = {2020},
  doi          = {10.3847/2041-8213/abbfad}
}

@article{RasioFord1996,
  author       = {Rasio, Frederic A. and Ford, Eric B.},
  title        = {Dynamical instabilities and the formation of extrasolar planetary systems},
  journal      = {Science},
  volume       = {274},
  pages        = {954--956},
  year         = {1996},
  doi          = {10.1126/science.274.5289.954}
}

@article{Chatterjee2008,
  author       = {Chatterjee, Sourav and Ford, Eric B. and Matsumura, Soko and Rasio, Frederic A.},
  title        = {Dynamical outcomes of planet-planet scattering},
  journal      = {ApJ},
  volume       = {686},
  pages        = {580--602},
  year         = {2008},
  doi          = {10.1086/590227}
}

@article{SutherlandFabrycky2016,
  author       = {Sutherland, Adam P. and Fabrycky, Daniel C.},
  title        = {On the Fate of Unstable Circumbinary Planets: Tatooine's Close Encounters with a Death Star},
  journal      = {ApJ},
  volume       = {818},
  pages        = {6},
  year         = {2016},
  doi          = {10.3847/0004-637X/818/1/6}
}

@article{Veras2016,
  author       = {Veras, Dimitri},
  title        = {Post-main-sequence planetary system evolution},
  journal      = {Royal Society Open Science},
  volume       = {3},
  number       = {6},
  pages        = {150571},
  year         = {2016},
  doi          = {10.1098/rsos.150571}
}

@book{BinneyTremaine2008,
  author       = {Binney, James and Tremaine, Scott},
  title        = {Galactic Dynamics},
  edition      = {2nd},
  publisher    = {Princeton University Press},
  year         = {2008}
}

@article{Paczynski1986,
  author       = {Paczynski, Bohdan},
  title        = {Gravitational microlensing by the galactic halo},
  journal      = {ApJ},
  volume       = {304},
  pages        = {1--5},
  year         = {1986},
  doi          = {10.1086/164140}
}

@article{Penny2019,
  author       = {Penny, Matthew T. and others},
  title        = {Predictions of the WFIRST Microlensing Survey. I. Bound Planet Detection Rates},
  journal      = {ApJS},
  volume       = {241},
  pages        = {3},
  year         = {2019},
  doi          = {10.3847/1538-4365/aafb69}
}

@article{Johnson2020,
  author       = {Johnson, Samuel A. and Penny, Matthew T. and Gaudi, B. Scott and others},
  title        = {Predictions of the Nancy Grace Roman Space Telescope Galactic Exoplanet Survey. II. Free-floating Planet Detection Rates},
  journal      = {AJ},
  volume       = {160},
  number       = {3},
  pages        = {123},
  year         = {2020},
  doi          = {10.3847/1538-3881/aba75b}
}

@article{McMillan_2016,
   title={The mass distribution and gravitational potential of the Milky Way},
   volume={465},
   ISSN={1365-2966},
   url={http://dx.doi.org/10.1093/mnras/stw2759},
   DOI={10.1093/mnras/stw2759},
   number={1},
   journal={Monthly Notices of the Royal Astronomical Society},
   publisher={Oxford University Press (OUP)},
   author={McMillan, Paul J.},
   year={2016},
   month=oct, pages={76–94} }

@ARTICLE{2024MNRAS.527..414C,
       author = {{Coleman}, Gavin A.~L. and {Nelson}, Richard P. and {Triaud}, Amaury H.~M.~J. and {Standing}, Matthew R.},
        title = "{Constraining the formation history of the TOI-1338/BEBOP-1 circumbinary planetary system}",
      journal = {\mnras},
         year = 2024,
        month = jan,
       volume = {527},
       number = {1},
        pages = {414-427},
          doi = {10.1093/mnras/stad3216},
archivePrefix = {arXiv},
       eprint = {2310.11898},
 primaryClass = {astro-ph.EP},
       adsurl = {https://ui.adsabs.harvard.edu/abs/2024MNRAS.527..414C}
}

@ARTICLE{2025arXiv251204700A,
       author = {{Albrow}, Leah and {Bannister}, Michele T. and {Forbes}, John C. and {Nesvorn{\'y}}, David},
        title = "{The ejection velocities of interstellar objects signpost their progenitor system architectures}",
      journal = {arXiv e-prints},
         year = 2025,
        month = dec,
          eid = {arXiv:2512.04700},
        pages = {arXiv:2512.04700},
          doi = {10.48550/arXiv.2512.04700},
archivePrefix = {arXiv},
       eprint = {2512.04700},
 primaryClass = {astro-ph.EP},
       adsurl = {https://ui.adsabs.harvard.edu/abs/2025arXiv251204700A}
}

@ARTICLE{2015MNRAS.449.3543W,
       author = {{Wang}, Long and {Kouwenhoven}, M.~B.~N. and {Zheng}, Xiaochen and {Church}, Ross P. and {Davies}, Melvyn B.},
        title = "{Close encounters involving free-floating planets in star clusters}",
      journal = {\mnras},
         year = 2015,
        month = jun,
       volume = {449},
       number = {4},
        pages = {3543-3558},
          doi = {10.1093/mnras/stv542},
archivePrefix = {arXiv},
       eprint = {1503.03077},
 primaryClass = {astro-ph.EP},
       adsurl = {https://ui.adsabs.harvard.edu/abs/2015MNRAS.449.3543W}
}

@ARTICLE{2022NatAs...6...89M,
       author = {{Miret-Roig}, N{\'u}ria and {Bouy}, Herv{\'e} and {Raymond}, Sean N. and {Tamura}, Motohide and {Bertin}, Emmanuel and {Barrado}, David and {Olivares}, Javier and {Galli}, Phillip A.~B. and {Cuillandre}, Jean-Charles and {Sarro}, Luis Manuel and {Berihuete}, Angel and {Hu{\'e}lamo}, Nuria},
        title = "{A rich population of free-floating planets in the Upper Scorpius young stellar association}",
      journal = {Nature Astronomy},
         year = 2022,
        month = feb,
       volume = {6},
        pages = {89-97},
          doi = {10.1038/s41550-021-01513-x},
archivePrefix = {arXiv},
       eprint = {2112.11999},
 primaryClass = {astro-ph.EP},
       adsurl = {https://ui.adsabs.harvard.edu/abs/2022NatAs...6...89M}
}

@misc{guo2025formationfreefloatingplanetsejection,
      title={Formation of Free-Floating Planets via Ejection: Population Synthesis with a Realistic IMF and Comparison to Microlensing Observations}, 
      author={Kangrou Guo and Shigeru Ida and Masahiro Ogihara},
      year={2025},
      eprint={2511.03246},
      archivePrefix={arXiv},
      primaryClass={astro-ph.EP},
      url={https://arxiv.org/abs/2511.03246}, 
}

@ARTICLE{2016MNRAS.461L.107M,
       author = {{Ma}, Sizheng and {Mao}, Shude and {Ida}, Shigeru and {Zhu}, Wei and {Lin}, Douglas N.~C.},
        title = "{Free-floating planets from core accretion theory: microlensing predictions}",
      journal = {\mnras},
         year = 2016,
        month = sep,
       volume = {461},
       number = {1},
        pages = {L107-L111},
          doi = {10.1093/mnrasl/slw110},
archivePrefix = {arXiv},
       eprint = {1605.08556},
 primaryClass = {astro-ph.EP},
       adsurl = {https://ui.adsabs.harvard.edu/abs/2016MNRAS.461L.107M}
}

@article{Chachan_2024,
   title={Planet Mass Function around M Stars at 1–10 au: A Plethora of Sub-Earth Mass Objects},
   volume={977},
   ISSN={1538-4357},
   url={http://dx.doi.org/10.3847/1538-4357/ad8c44},
   DOI={10.3847/1538-4357/ad8c44},
   number={1},
   journal={The Astrophysical Journal},
   publisher={American Astronomical Society},
   author={Chachan, Yayaati and Lee, Eve J.},
   year={2024},
   month=dec, pages={61} }

@article{Cumming2008,
  author  = {Cumming, Andrew and Butler, R. Paul and Marcy, Geoffrey W. and Vogt, Steven S. and Wright, Jason T. and Fischer, Debra A.},
  title   = {The Keck Planet Search: Detectability and the Minimum Mass and Orbital Period Distribution of Extrasolar Planets},
  journal = {Publications of the Astronomical Society of the Pacific},
  volume  = {120},
  number  = {867},
  pages   = {531--554},
  year    = {2008},
  doi     = {10.1086/588487}
}

@article{Fernandes2019,
  author       = {Fernandes, R. B. and Mulders, G. D. and Pascucci, I. and Mordasini, C. and Emsenhuber, A.},
  title        = {Hints for a Turnover at the Snow Line in the Giant Planet Occurrence Rate},
  journal      = {The Astrophysical Journal},
  volume       = {874},
  number       = {1},
  pages        = {81},
  year         = {2019},
  doi          = {10.3847/1538-4357/ab0300},
  eprint       = {1812.05569},
  archivePrefix= {arXiv},
  primaryClass = {astro-ph.EP}
}

@article{Armstrong2014,
  author       = {Armstrong, D. J. and Osborn, H. P. and Brown, D. J. A. and Faedi, F. and Gomez Maqueo Chew, Y. and Martin, D. V. and Pollacco, D. and Udry, S.},
  title        = {On the Abundance of Circumbinary Planets},
  journal      = {Monthly Notices of the Royal Astronomical Society},
  volume       = {444},
  number       = {3},
  pages        = {1873--1883},
  year         = {2014},
  doi          = {10.1093/mnras/stu1570},
  eprint       = {1404.5617},
  archivePrefix= {arXiv},
  primaryClass = {astro-ph.EP}
}

@article{Coleman2024,
    author  = {Coleman, Gavin A. L.},
    title   = {On the properties of free-floating planets originating in circumbinary planetary systems},
    journal = {Monthly Notices of the Royal Astronomical Society},
    volume  = {530},
    number  = {1},
    pages   = {630--644},
    year    = {2024},
    doi     = {10.1093/mnras/stae903}
}

@article{ColemanDeRocco2025,
    author  = {Coleman, Gavin A. L. and DeRocco, William},
    title   = {Predicting the Galactic population of free-floating planets from realistic initial conditions},
    journal = {Monthly Notices of the Royal Astronomical Society},
    volume  = {537},
    number  = {3},
    pages   = {2303--2312},
    year    = {2025},
    doi     = {10.1093/mnras/staf138}
}

@article{HuangLai2026,
    author  = {Huang, Xiumin and Lai, Dong},
    title   = {Free-floating Planets Produced by Planet--Planet Scatterings: Ejection Velocity and Survival Rate of Their Moons},
    journal = {The Astrophysical Journal},
    volume  = {998},
    number  = {2},
    pages   = {245},
    year    = {2026},
    doi     = {10.3847/1538-4357/ae394f},
    eprint  = {2508.18239},
    archivePrefix = {arXiv},
    primaryClass  = {astro-ph.EP}
}

@article{Piffl2014,
    author = {Piffl, T. and Scannapieco, C. and Binney, J. and
              Steinmetz, M. and Scholz, R.-D. and Williams, M. E. K. and
              de Jong, R. S. and Kordopatis, G. and Matijevi{\v c}, G. and
              Bienaym{\'e}, O. and Bland-Hawthorn, J. and Boeche, C. and
              Freeman, K. and Gibson, B. and Gilmore, G. and
              Grebel, E. K. and Helmi, A. and Munari, U. and
              Navarro, J. F. and Parker, Q. and Reid, W. A. and
              Seabroke, G. and Watson, F. and Wyse, R. F. G. and
              Zwitter, T.},
    title   = {The {RAVE} survey: the Galactic escape speed and the mass of the Milky Way},
    journal = {Astronomy \& Astrophysics},
    volume  = {562},
    pages   = {A91},
    year    = {2014},
    doi     = {10.1051/0004-6361/201322531},
    eprint  = {1309.4293},
    archivePrefix = {arXiv},
    primaryClass  = {astro-ph.GA}
}

@article{Monari2018,
    author = {Monari, Giacomo and Famaey, Benoit and Carrillo, Ismael and
              Piffl, Tilmann and Steinmetz, Matthias and Wyse, Rosemary F. G. and
              Anders, Friedrich and Chiappini, Cristina and Jan{\ss}en, Katja},
    title   = {The escape speed curve of the Galaxy obtained from {Gaia DR2} implies a heavy Milky Way},
    journal = {Astronomy \& Astrophysics},
    volume  = {616},
    pages   = {L9},
    year    = {2018},
    doi     = {10.1051/0004-6361/201833748},
    eprint  = {1807.04565},
    archivePrefix = {arXiv},
    primaryClass  = {astro-ph.GA}
}

@article{Deason2019,
    author = {Deason, Alis J. and Fattahi, Azadeh and Belokurov, Vasily and
              Evans, N. Wyn and Grand, Robert J. J. and Marinacci, Federico and
              Pakmor, R{\"u}diger},
    title   = {The local high-velocity tail and the Galactic escape speed},
    journal = {Monthly Notices of the Royal Astronomical Society},
    volume  = {485},
    number  = {3},
    pages   = {3514--3526},
    year    = {2019},
    doi     = {10.1093/mnras/stz623},
    eprint  = {1901.02016},
    archivePrefix = {arXiv},
    primaryClass  = {astro-ph.GA}
}

@article{Dehnen:1999ea,
    author = "Dehnen, Walter",
    title = "{Simple distribution functions for stellar disks}",
    eprint = "astro-ph/9906082",
    archivePrefix = "arXiv",
    reportNumber = "OUTP99-08A",
    doi = "10.1086/301010",
    journal = "Astron. J.",
    volume = "118",
    pages = "1201",
    year = "1999"
}

@ARTICLE{2001MNRAS.322..859B,
       author = {{Bonnell}, Ian A. and {Smith}, Kester W. and {Davies}, Melvyn B. and {Horne}, Keith},
        title = "{Planetary dynamics in stellar clusters}",
      journal = {\mnras},
         year = 2001,
        month = apr,
       volume = {322},
       number = {4},
        pages = {859-865},
          doi = {10.1046/j.1365-8711.2001.04171.x},
archivePrefix = {arXiv},
       eprint = {astro-ph/0012020},
 primaryClass = {astro-ph},
       adsurl = {https://ui.adsabs.harvard.edu/abs/2001MNRAS.322..859B}
}

@ARTICLE{2026PASP..138h2001R,
       author = {{Rice}, Malena and {DeRocco}, William and {Raymond}, Sean N.},
        title = "{The Dynamics of Planetary Ejection}",
      journal = {\pasp},
         year = 2026,
        month = aug,
       volume = {138},
       number = {8},
          eid = {082001},
        pages = {082001},
          doi = {10.1088/1538-3873/ae9343},
archivePrefix = {arXiv},
       eprint = {2608.00173},
 primaryClass = {astro-ph.EP},
       adsurl = {https://ui.adsabs.harvard.edu/abs/2026PASP..138h2001R}
}

@misc{albrow2025ejectionvelocitiesinterstellarobjects,
      title={The ejection velocities of interstellar objects signpost their progenitor system architectures}, 
      author={Leah Albrow and Michele T. Bannister and John C. Forbes and David Nesvorný},
      year={2025},
      eprint={2512.04700},
      archivePrefix={arXiv},
      primaryClass={astro-ph.EP},
      url={https://arxiv.org/abs/2512.04700}, 
}

@misc{boffin2024importancebinarystars,
      title={The importance of binary stars}, 
      author={Henri M. J. Boffin and David Jones},
      year={2024},
      eprint={2411.18470},
      archivePrefix={arXiv},
      primaryClass={astro-ph.SR},
      url={https://arxiv.org/abs/2411.18470}, 
}

\end{document}